\documentclass[fleqn,usenatbib]{mnras}

\usepackage{newtxtext,newtxmath}

\usepackage[T1]{fontenc}
\usepackage{lipsum}
\usepackage{subcaption}

\DeclareRobustCommand{\VAN}[3]{#2}
\let\VANthebibliography\thebibliography
\def\thebibliography{\DeclareRobustCommand{\VAN}[3]{##3}\VANthebibliography}

\usepackage{graphicx}	
\usepackage{amsmath}	

\title[Optimized \texttt{HDBSCAN} for MW substructure]{Optimized \texttt{HDBSCAN} clustering for reconstructing the merger history of the Milky Way: applications and limitations}

\author[A. Sante et al.]{
Andrea Sante,$^{1}$\thanks{E-mail: A.Sante@2022.ljmu.ac.uk}
Andreea S. Font,$^{1}$
Dharmesh Mistry,$^{2}$
Sandra Ortega-Martorell,$^{2}$
and Ivan Olier$^{2}$
\\
$^{1}$Astrophysics Research Institute, Liverpool John Moores University, 146 Brownlow Hill, Liverpool L3 5RF, UK\\
$^{2}$School of Computer Science and Mathematics, Liverpool John Moores University, James Parsons Building, 3 Byrom Street, Liverpool, L3 3AF, UK
}

\date{Accepted XXX. Received YYY; in original form ZZZ}

\pubyear{\the\year{}}

\begin{document}
\label{firstpage}
\pagerange{\pageref{firstpage}--\pageref{lastpage}}
\maketitle

\begin{abstract}
Clustering algorithms can help reconstruct the assembly history of the Milky Way by identifying groups of stars sharing
similar properties in a kinematical or chemical abundance space. Despite being promising tools, their efficiency has
not yet been fully tested in a realistic cosmological framework. 
We investigate the effectiveness of the HDBSCAN clustering algorithm in the recovery of the progenitors of Milky Way-type galaxies, using several systems from the Auriga suite of simulations. We develop a methodology aimed at improving the efficiency of the algorithm and avoiding fragmentation: First, we use a 12-dimensional feature space including a range of chemodynamical properties and stellar ages; furthermore, we optimise the algorithm using information from the internal structure of the clusters of accreted stars. We show that our approach yields good results in terms of both purity and completeness of clusters for galaxies with different types of accretion histories. We also evaluate the decrease in efficiency due to contamination by in situ stars. While for accreted-only haloes the algorithm matches well the recovered clusters with the individual progenitors and is able to recover accretion events up to a redshift of accretion $z_{\rm acc}\sim3$, for accreted + in situ haloes it can only identify the more recent accretion events ($z_{\rm acc} < 1$). However, the purity of the identified clusters remains remarkably high even in this case. Our results suggest that HDBSCAN can efficiently identify accreted debris in Milky Way-type galaxies in realistic conditions, however, it requires careful optimization to provide valid results.
\end{abstract}

\begin{keywords}
 Galaxy: halo -- Galaxy: stellar content --  software: machine learning -- software: simulations
\end{keywords}



\section{Introduction}

In a $\Lambda$CDM cosmological model, galaxies similar to the Milky Way grow hierarchically, by accretion and disruption of a multitude of satellite dwarf galaxies \citep{Searle1978,White1978}. The remnants of the disruption events are expected to reside in the stellar halo, in the form of tidal streams of various degrees of phase-mixing \citep{Helmi2020}. The identification of past merger events can be done, in principle, by using a information retained in their kinematics using the assumption that certain kinematical properties are conserved, i.e., energy, angular momenta, actions \citep{Helmi_1999b,Helmi2000,Knebe2005,Font_2006a,Morrison2009,Gomez2010}. Moreover, stars from a specific progenitor are also expected to retain a memory of their common star formation history and, implicitly, chemical evolution (\citealt{FreemanBlandHawthorn_2002}; see also the reviews of \citealt{BlandHawthorn2016} and \citealt{Deason2024}). 

The increased availability of high-quality astrometric data from Gaia \citep{Gaia2016,GaiaDR3} and complementary chemical abundances from spectroscopic surveys such as APOGEE \citep{Apogee}, Gaia-ESO \citep{Gilmore2012,Randich2013}, LAMOST \citep{lamost}, H3 \citep{Conroy2019}, or GALAH \citep{Buder2021,galah} has allowed the discovery of several substructures in the Galaxy. The debris identified so far appear as distinct features in the chemodynamical space \citep[e.g.,][]{Klement2009,Helmi2018,Belokurov2018,Koppelman_2019,Kruijssen2019,Ji2020,Naidu2020,Horta2023,Malhan2024}.

However, the identification of debris remains challenging in the presence of the background population and uncertainties introduced by instrumental effects \citep{Brown2005}, and given that overdensities originating from multiple progenitors overlap in kinematical space and share similar chemical abundances \citep{Jean-Baptiste2017,Orkney2023,Carrillo_2024,Mori2024,Kizhuprakkat2024}. 

Machine learning methods, and clustering algorithms in particular, are well-suited for identifying halo substructures, given the natural tendency of stars from disrupted progenitors to cluster in kinematical space or display similar chemical abundances, in correspondence to the properties of progenitor galaxies they originate from. Indeed, clustering methods have shown their usefulness by detecting several new stellar substructures in the Milky Way and confirming some of the existing ones \citep[e.g.,][]{Koppelman_2019,Myeong2019,Borsato2020,Necib2020,ReFiorentin2015,Ruiz-Lara2022,Yuan2020,Dodd2023,Kim2025}. 

However, clustering methods also have some limitations when applied to Galactic archaeology. One example is the family of partitioning algorithms, such as $k$-means \citep{Kmeans}, which aim to divide the data into distinct groups by minimizing the intra-cluster variance. Their requirement to predefine the number of clusters makes them less ideal for identifying debris from Milky Way progenitors, where this number is not known {\it a priori}. These methods are more useful when the clusters correspond to known structures, such as multiple stellar populations of globular clusters \citep{Fraix-Burnet2015}, or the distinct Galactic components \citep{Mackereth2019}. Moreover, the $k$-means algorithm is biased toward identifying spherically clustered groups of stars, which makes it unsuited for picking out tidal debris that typically displays an asymmetric distribution. For example, instead of ``clumps'' in the energy -- angular momentum space,  as expected from conservation of these two quantities for any progenitor, the overdensities in this space could be severely distorted as debris orbits in an evolving gravitational potential or interacts with other substructures in the host galaxy \citep[e.g.,][]{Font_2006a,Gomez2010,Gomez2013}.

Hierarchical clustering methods \citep[see review of][]{hierarchical_clustering}, in particular the so-called single-linkage methods, build a nested tree of clusters by progressively merging the two closest clusters in the parameter space (or feature space, in machine learning terminology) until every point belongs to the same cluster. Despite mimicking the hierarchical nature of galaxy formation, these methods incorporate every data point in a hierarchy which makes them sensitive to noise. Isolated points in the feature space can act as a bridge between clusters or distort the internal structure of clusters. Moreover, extracting a final, ``flat'' partition from the hierarchy is often not straightforward, as meaningful clusters, representing coherent debris from distinct accretion events, can appear at various levels within the hierarchy, reflecting differences in their spatial extent and detectability. For example, a tightly clustered debris from a recently accreted progenitor might remain distinct at a higher level in the hierarchy, whereas a more dispersed debris from an early accretion might merge with the background noise at a lower level. Using the single linkage-based clustering methodology developed by \citet{Lovdal_2022} on the \texttt{Auriga} simulations, \citet{Thomas2025} found that only recently accreted ($<6-7$~Gyr ago) progenitors are efficiently identified. They also found that the method is prone to generate artificial detections, with a significant fraction of clusters being composed mainly of in situ stars.

Density-based clustering methods \citep[e.g., DBSCAN, HDBSCAN;][]{dbscan, optics, Campello_hdbscan_2013} explicitly account for the presence of noise points by defining clusters as dense regions of points separated by sparser regions of noise. These methods are well suited for the distribution of debris from individual progenitor galaxies, which correspond to overdensities in the feature space against a background of unrelated in situ stars. Furthermore, density-based methods make no assumptions about the shape of a cluster, allowing for the identification of elongated filamentary structures characteristic of tidal streams.  \texttt{HDBSCAN} \citep[Hierarchical Density-Based Spatial Clustering of Applications with Noise,][]{Campello_hdbscan_2013} offers significant advantages over its predecessor, \texttt{DBSCAN} \citep{dbscan}. Although DBSCAN relies on a single distance scale to define dense regions, \texttt{HDBSCAN} identifies clusters at various scales simultaneously, providing a hierarchical representation of structures in the feature space. This is important as different progenitor systems can have specific internal velocity dispersions or metallicity distributions, which result in structures of different overdensities in the feature space.
Moreover, \texttt{HDBSCAN} automates the extraction of a flat partition by selecting the most persistent and significant clusters in the hierarchy based on a cluster stability metric.

However, as with other clustering algorithms, the configuration of the clusters retrieved by \texttt{HDBSCAN} varies considerably with the choice of its internal parameters. An improper parameter selection can lead to either the fragmentation of larger, coherent structures or the erroneous aggregation of unrelated stars into spurious clumps \citep{Kim2025}. 

In this study, we evaluate the effectiveness of \texttt{HDBSCAN} on simulated stellar haloes, where the ``ground truth'' is known (i.e., tidal debris is already identified and mapped to its corresponding progenitor galaxies, and contamination from in situ stars is also known). We use a sample of three Milky Way-mass galaxies from \texttt{Auriga} suite \citep{Grand_2017,Grand_2024}, which has been shown to form realistic disk-dominated galaxies similar to the Milky Way \citep{Grand_2017, Orkney2022}. The three galaxies are chosen so that they have significantly different assembly histories, in order to evaluate the variations in the performance of \texttt{HDBSCAN} with different chemodynamical distributions of tidal debris. We also incorporate a wider range of chemodynamical properties (features) in the analysis than previously used in similar clustering algorithms.

To avoid cluster fragmentation, we implement a systematic optimization of the \texttt{HDBSCAN} parameters tested against various internal and external metrics. The internal metrics account for how the data are partitioned by the algorithm, while the external metrics take into account the similarity between the clusters and the actual progenitor groups. Detailed explanations of the \texttt{HDBSCAN} algorithm, the optimization of its parameters, and the metrics used to evaluate its effectiveness are provided in Section~\ref{sec:hdbscan_algorithm}.

Furthermore, the stellar halo of the Milky Way is known to have two distinct components, accreted and formed in situ \citep{Carollo2007,Beers2012,Belokurov2022}. Cosmological simulations naturally produce dual haloes \citep{Zolotov_2009,Font2011,McCarthy2012,Tissera2013,Cooper2015}. Because the two halo populations can have overlapping chemodynamical properties \citep{Haywood2018, DiMatteo2019, Mori2024}, the performance of the clustering algorithm generally decreases in the presence of in situ stars or can lead to the identification of spurious substructures. To rigorously validate the efficacy of our optimized \texttt{HDBSCAN} methodology, we first apply it to accreted-only components. This allows us to quantify the intrinsic ability of the algorithm in recovering accreted structures of varying morphologies and densities without the contamination from the in situ population, hence establishing a baseline for interpreting the results. With this baseline established, we then apply the algorithm to the fully mixed stellar halo.  

The paper is structured as follows. Section~\ref{sec:data} describes the \texttt{Auriga} suite of simulations and the selection of our sample of galaxies, including a brief overview of their distinct assembly histories.
Section~\ref{sec:results} presents the results of the optimized \texttt{HDBSCAN}, separately for the accreted-only (Sec.~\ref{sec:accreted_only}) and accreted + in situ stellar haloes (Sec.~\ref{sec:with_in situ}). The implications of our findings for reconstructing the accretion history of the Milky Way and the potential future directions of research are discussed in Section~\ref{sec:discussions}.

\section{The \texttt{Auriga} Simulations}
\label{sec:data}

The \texttt{Auriga} suite comprises 39 magnetohydrodynamical (MHD) simulations of Milky Way-mass ($0.5<M_{200}/(10^{12} \, \mathrm{M}_{\odot})<2$) haloes run with the \texttt{AREPO} \citep{Springel_2010} moving-mesh code at two resolution levels. In the following, we provide a brief description of the \texttt{Auriga} suite and refer the reader to \citet{Grand_2017} and \citet{Grand_2024} for more details.

The Milky Way-mass haloes were randomly selected from the most isolated quartile of their respective mass ranges at $z=0$ in the parent L100N1504 \texttt{EAGLE} dark-matter only simulation \citep{Schaye_eagle_2015}.  Zoomed-in simulations were then performed, following the particles within a distance of $4\, R_{200}$\footnote{Galactocentric distance which defines a spherical region where the average matter density is $200$ times the critical matter density of the universe. It is often used as a proxy for the virial radius of a halo.} from the center of the haloes to $z=127$ and sampling lower mass particles in the enclosed area with higher mass particles added at larger distances. A $\Lambda$ cold dark matter cosmology with parameters taken from Planck Collaboration XVI (\citeyear{Planck_collab_2014}) was adopted. Baryons were introduced by splitting the particles into dark matter particle-gas cell pairs according to the cosmological baryon mass fraction. The evolution of the MHD equations in a cosmological context was followed by \texttt{AREPO} through a discretization of the simulation domain carried out with the Voronoi tessellation of mesh-generating points that move with the velocity of the fluid flow. Each cell was assigned a given mass so that high-density regions are resolved with a larger number of cells.

The \texttt{Auriga} galaxy formation model incorporates a wide range of subgrid physics, including primordial and metal line cooling with self-shielding corrections against a spatially uniform, redshift-dependent UV background field \citep{Faucher-Giguere_2009} that completes reionization at $z\sim6$ \citep{Vogelsberger_2013}. The interstellar medium (ISM) is treated using the two-phase model of \cite{Springel_ISM_2003}, with stochastic star formation occurring in gas denser than $n=0.13\,\mathrm{cm}^{-3}$ assuming a \cite{Chabrier_imf_2003} initial mass function. The model also accounts for stellar evolution, tracking the associated mass loss and chemical enrichment of the surrounding gas from asymptotic giant branch stars \citep{Karakas_2010}, Type Ia supernovae \citep{Thielemann2003, travaglio2004} and Type II supernovae \citep{Portinari1998}. Feedback is implemented through stellar wind particles released from Type II supernovae, as well as with radiative and thermal energy outbursts from active galactic nuclei with both radio and quasar modes.
The model also accounts for the seeding and growth of supermassive black holes \citep{Springel2005} and includes a uniform, comoving primordial magnetic field of $10^{-14}\,\mathrm{G}$ \citep{pakmor_2013}.

To trace the formation history of galaxies, simulation outputs (``snapshots'') were post-processed to identify structures and link them through time. First, Friends-of-Friends (FOF) groups were identified based on the dark matter particle spatial distribution. Within these FOF groups, the \texttt{SUBFIND} algorithm \citep{Springel_2001} was run on all particle types to identify gravitationally self-bound substructures, or subhaloes. These subhaloes were then linked across snapshots to construct merger trees. In this process, a subhalo in a given snapshot was identified as the main progenitor of a subhalo in a subsequent snapshot if it contributes the highest number of particles, weighted by their binding energy. 

Star particles were tagged as either accreted or formed in situ, by tracing them back to their point of origin: accreted if they formed within a satellite dwarf galaxy (either before or after accretion onto the host) and are gravitationally bound to the main halo at $z=0$, and in situ if they formed in the main galaxy \citep{Monachesi_auriga_2019}. We use this information here, together with additional data on the accreted star particles available in the \texttt{Auriga} public database\footnote{This information is provided for each simulation as part of the ``Accreted particle lists'' catalogue downloadable at \url{https://wwwmpa.mpa-garching.mpg.de/auriga/data_new.html}.}, namely the merger tree index of the subhalo where each accreted star particle formed, \texttt{RootIndex}, and the index of the subhalo to which the star particles were bound at the time when the subhalo reached its maximum stellar mass, \texttt{PeakMassIndex} (see \citealt{Grand_2024}). 

In our analysis, a progenitor of a Milky Way-mass galaxy is defined as a group of accreted stars\footnote{We note that the term ``stars" in the context of simulations refers to star particles, which are essentially single stellar populations.} that share the same \texttt{PeakMassIndex}; henceforth, we will refer to these progenitors based on their respective numerical index, i.e., ``prog.~\texttt{PeakMassIndex}''. We motivate this definition by the fact that subhaloes tend to reach their maximum stellar mass just before accretion onto their host, that is, when they first cross the $R_{200}$ of their host. At around this time, which we denote $\tau_{\rm infall}$, most infalling satellites experience tidal and ram-pressure stripping, which results in a sharp decrease in their star formation (e.g., \citealt{Kawata_2008, Simpson_2018, Font_quenching_2022}). We therefore evaluate the performance of the \texttt{HDBSCAN} algorithm by comparing the clusters of accreted stars identified by the algorithm with the \texttt{PeakMassIndex}-defined progenitor groups.

The Milky Way-mass galaxies in \texttt{Auriga} match the galaxy scaling relations, e.g., the stellar mass-halo mass and stellar mass-metallicity relations \citep{Grand_2017}. They also have stellar haloes with similar stellar masses, surface brightnesses and metallicity profiles of observed Milky Way analogues \citep{Monachesi_auriga_2019}. Morphologically, they are also similar to the Milky Way at present time (i.e., they have rotationally supported discs). In particular, for our study, the suite of simulations includes a variety of assembly histories, which allows us to test the effectiveness of the \texttt{HDBSCAN} algorithm on different chrono-chemodynamical distributions of stellar haloes.

\subsection{The simulations sample}
\label{sec:sims_sample}

The three galaxies considered here, Au 7, Au 25 and Au 27, are selected from the original sample of $30$ Milky Way-mass systems in the ``level~4'' suite of simulations. The particle mass resolution is $\sim 4 \times 10^{5}\, \mathrm{M}_{\odot}$ for dark matter and, typically, $\sim 5 \times 10^{4} \, \mathrm{M}_{\odot}$ for baryons.  

The three host galaxies display significantly different present-day distributions of accreted stars, in either kinematics, chemical abundances or ages. Au 7 has a predominantly early formation epoch, and consequently the accreted debris is mostly phase-mixed at $z=0$; in contrast, Au 25 has a late formation, and its debris is mostly in the form of cold stellar streams; Au27 has an assembly history more similar to the Milky Way, including an early major merger similar to the Gaia Enceladus-Sausage (GES) \citep{Fattahi_2019}. Au 27 also has a large stellar disc and a bar at $z=0$. 

In addition to the information from the merger trees, we also use the morphological classification of tidal debris in this \texttt{Auriga} suite performed by \cite{riley_2025} and \cite{shipp_2025}. This classification groups accretion events into three categories, ``intact'', ``phase-mixed'' and ``stellar streams'', based on their configuration at $z=0$.  An accretion event is classified as intact if the majority ($>97\%$) of accreted stars are still bound to the system at present. The debris from disrupted satellites is considered either phase-mixed or as (cold) stellar stream based on a mass-dependent cut in the velocity dispersion, with the phase-mixed structures exhibiting a higher velocity dispersion than the streams.

For each galaxy, we consider only the accreted stars that contribute to the main halo, defined as those accretion events that contribute at least 0.1\% to the total number of accreted particles at $z=0$. This corresponds to a cutoff in the number of particles per progenitor ranging between 100 and 1000, depending on the system. We also exclude particles that are bound to surviving satellites and those from merger events labeled as ``intact'',  under the assumption that the identification of these systems would not require the application of clustering techniques.

The assembly histories of the selected galaxies are briefly summarized below, with an overview of the number and properties of their most significant progenitors given in Table~\ref{tab:galaxy_sample}. The merger trees for the galaxies in the sample are shown in the first row of Fig.~\ref{fig:merger_tree_acc}.

\subsubsection{Au7}

The majority ($>90\%$) of the accreted stellar content in Au7 is in the form of phase-mixed structures. These structures originate mostly from three progenitors (prog.~\texttt{3210}, prog.~\texttt{4401} and prog.~\texttt{2197}) of similar stellar mass ($\sim10^{10}~\mathrm{M_{\sun}}$), which contribute 20\%, 34\% and 36\%, respectively, to the accreted halo mass. These events succeeded each other almost regularly every $\sim3~\mathrm{Gyr}$ starting from $z\sim1$.

\subsubsection{Au25}

Most of the accreted star particles bound to the main halo at $z=0$ in Au25 originate from a satellite of stellar mass $\sim10^{10}~\mathrm{M_{\sun}}$, prog.~\texttt{736}, accreted onto the main halo $<2~\mathrm{Gyr}$ ago. However, these comprise only a small fraction ($\sim5\%$) of the stellar content of the satellite, which is still present as a self-bound system at $z=0$. The other massive accretion event experienced by the Au25 was a satellite of stellar mass $\sim10^{8.5}~\mathrm{M_{\sun}}$, prog.~\texttt{41397}, accreted at $z_{\rm acc}>2$. The stars disrupted from this progenitor form a phase-mixed structure at present, comprising $\sim34\%$ of the accreted stars of its host.

\subsubsection{Au27}

The accreted halo in Au27 is dominated by stars originating from a massive satellite (stellar mass of $\sim10^{9.5}~\mathrm{M_{\sun}}$),  prog.~\texttt{7219}, accreted on a radial orbit at $z_{\rm acc}\sim1.5$, similar to a GES event \citep{Helmi2018,Belokurov2018}. This contributed $\sim40\%$ of the accreted star particles bound to the main halo at $z=0$. Two other significant, slightly less massive satellites were accreted just before, contributing 19\% and, respectively, 14\% to the mass of the accreted halo. Although most of the accreted debris are completely mixed in with in situ stars, 13\% of the accreted stars still present stellar stream-like configurations; such an example is the remnant of the massive satellite,  prog.~\texttt{205}, accreted $\sim 7~\mathrm{Gyr}$ ago. Au27 also presents a companion of similar mass as the Large Magellanic Cloud at $z=0$, which, however, is not considered in the analysis.

\begin{table}
    \centering
    \begin{tabular}{l|c|c|c|c}
        \hline
        \texttt{PeakMassIndex} & $\tau_{\mathrm{infall}} \; [\mathrm{Gyr}]$ & $\log(M_*/\mathrm{M}_{\odot})$ & $f_{\mathrm{acc}}$ &
        Morphology\\
        \hline
        \multicolumn{5}{c}{{\bf Au7}} \\
        prog.~\texttt{2197} & 2.8 & 9.8 & 0.36 & phase-mixed \\
        prog.~\texttt{2185} & 4.6 & 8.7 & 0.02 & stream \\
        prog.~\texttt{4401} & 5.9 & 9.7 & 0.34 & phase-mixed \\
        prog.~\texttt{3210} & 8.4 & 9.6 & 0.20 & phase-mixed \\
        prog.~\texttt{3059} & 8.4 & 8.5 & 0.02 & stream \\
        prog.~\texttt{157} & 10.2 & 8.9 & 0.04 & phase-mixed \\
        \hline
        \multicolumn{5}{c}{{\bf Au25}} \\
        prog.~\texttt{736} & 1.7 & 10.2 & 0.44 & stream \\
        prog.~\texttt{4643} & 6.2 & 8.1 & 0.08 & stream \\
        prog.~\texttt{692} & 9.3 & 7.6 & 0.03 & stream \\
        prog.~\texttt{19027} & 10.7 & 8.1 & 0.07 & phase-mixed \\
        prog.~\texttt{41397} & 11.3 & 8.7 & 0.34 & phase-mixed \\
        \hline
        \multicolumn{5}{c}{{\bf Au27}} \\
        prog.~\texttt{205} & 6.8 & 9.0 & 0.13 & phase-mixed \\
        prog.~\texttt{6355} & 6.8 & 8.0 & 0.01 & stream \\
        prog.~\texttt{4434} & 6.8 & 8.0 & 0.01 & stream \\
        prog.~\texttt{9073} & 7.3 & 8.8 & 0.05 & phase-mixed \\
        prog.~\texttt{7219} & 9.6 & 9.6 & 0.40 & phase-mixed \\
        prog.~\texttt{173} & 10.4 & 9.1 & 0.14 & phase-mixed \\
        prog.~\texttt{168} & 10.7 & 9.2 & 0.20 & phase-mixed \\
        prog.~\texttt{21670} & 11.0 & 8.6 & 0.03 & phase-mixed \\
        \hline
    \end{tabular}
    \caption{Merger events contributing at least 1\% of the accreted stars bound to the main halo at $z=0$ for the galaxies in the sample.  From the leftmost to the rightmost, the columns describe: (i) the index of the progenitor in the merger tree of the galaxy; (ii) the lookback infall time, i.e., the first time the progenitor crossed $R_{200}$ of its host; (iii) the stellar mass of the progenitor at the time of infall; (iv) the fraction of star particles bound to the main halo at $z=0$ which were accreted from the progenitor; (v) the present-day configuration of the progenitor debris as classified by \citet{Riley_aurigastreams1_2024}.}
    \label{tab:galaxy_sample}
\end{table}

\section{The \texttt{HDBSCAN} methodology}
\label{sec:hdbscan_algorithm}

\texttt{HDBSCAN} \citep{Campello_hdbscan_2013} is a non-parametric clustering algorithm that identifies clusters of varying shapes and densities, while also identifying data points that do not belong to any cluster, termed noise. It extends the \texttt{DBSCAN} algorithm by producing a hierarchical clustering structure from which a simplified set of flat clusters can be extracted. The methodology begins by transforming the feature space to reflect the local density of data points through the concept of \textit{core distance}. For each data point $x$, its core distance, $core_k(x)$, is the distance to its $k^{\rm th}$ nearest neighbor, providing an estimate of the local density.

A new distance metric, the \textit{mutual reachability distance},  is then defined between two points, $a$ and $b$, as
\begin{equation*}
   d_{mreach-k}(a,b) = \max(core_k(a), core_k(b), d(a,b)), 
\end{equation*}
where $d(a,b)$ is the original distance. This metric effectively smooths out the density of the data space by ensuring that points in dense regions remain close while pushing points in sparse regions further apart.

The data set is then treated as a fully connected graph, where the edge weights are the mutual reachability distances. From this graph, a \textit{minimum spanning tree} (MST) is constructed. This is a subgraph connecting all points with the minimum possible total edge weight and no cycles. The MST captures the underlying structure of the data.

Afterward, a cluster hierarchy is created by sorting the MST edges by weight in increasing order and iteratively merging the components connected by each edge. This process results in a tree of nested clusters. This full hierarchy is then simplified by a condensation process governed by the \texttt{min\_cluster\_size} parameter (see Section~\ref{sec:hdbscan_tuning}). Traversing the hierarchy from the root, if a split results in a new cluster with fewer points than \texttt{min\_cluster\_size}, that component is deemed to be made of points that have "fallen out of a cluster" and is not considered for further partitioning. The final step is to extract a stable, flat set of clusters from this condensed hierarchy. A stability score is calculated for each cluster based on its persistence or "lifespan" within the hierarchy. The clusters with the highest stability are selected as the final output, and any point not belonging to a selected cluster is classified as noise.

\subsection{Optimizing the \texttt{HDBSCAN} parameters with \texttt{OPTUNA}}
\label{sec:hdbscan_tuning}

For the analysis, we use the algorithm implementation provided by the \texttt{HDBSCAN} \texttt{Python} package \citep{hdbscan_software}. The functioning of a \texttt{HDBSCAN} clustering model is governed primarily by the following parameters:

\begin{itemize}
    \item \texttt{min\_samples}, which sets the minimum number of samples required in a neighbourhood for a point to be considered a core point. This parameter influences the core distance calculation and dictates how conservatively the density is estimated. A higher value results in larger core distances, increasing the likelihood that points in low-density regions are classified as noise.

    \item \texttt{min\_cluster\_size}, which defines the minimum number of points a cluster must have to be considered a genuine cluster during the hierarchy condensation step.

    \item \texttt{cluster\_selection\_epsilon}, which specifies a distance threshold for cluster extraction. When traversing the condensed hierarchy, if a split from a parent to its children occurs over a distance greater than \texttt{cluster\_selection\_epsilon}, the parent cluster is selected instead of its children. This allows for the extraction of clusters at a specific density level.
\end{itemize}

The choice of these parameters can affect the way the data are grouped into clusters, which consequently impacts the efficiency of \texttt{HDBSCAN} in retrieving the progenitors of a galaxy.  To ensure an objective and reproducible analysis, we have automated the parameter selection process. This process involves the definition of a search space for the relevant parameters, running \texttt{HDBSCAN} models for different configurations sampled from this space, and evaluating the resulting clusters with a given metric. We implemented this optimization using the \texttt{Optuna} hyperparameter search framework \citep{optuna_2019}. \texttt{Optuna} efficiently samples the parameter space by constructing a probabilistic model from past trials to predict which configurations are most likely to maximize the evaluation metric.

Specifically, the parameter ranges explored in the search are:
\begin{itemize}

\item \texttt{min\_samples} $\in [5, 50]$
\item \texttt{min\_cluster\_size} $\in [50, 500]$, and
\item \texttt{cluster\_selection\_epsilon} $\in [0.0, 1.0]$.
\end{itemize}

Two distinct evaluation metrics were considered to guide the optimization:
\begin{enumerate}
\item the V-measure \citep{vmeasure}, an external entropy-based metric that compares the particle IDs in the \texttt{HDBSCAN} clusters to the ones in the  progenitor groups defined by \texttt{PeakMassIndex}. It is defined by a combination of two criteria: homogeneity, which is satisfied if each cluster contains only particles of a single progenitor, and completeness, which is satisfied if all members of a given progenitor are assigned to the same cluster. The V-measure score ranges from 0 to 1, with higher scores indicating better clustering results. A score of 1 represents a perfectly homogeneous and complete clustering, meaning each cluster contains only particles from a single progenitor, and all particles of a given progenitor are grouped together in one cluster. A score of 0 indicates a completely random or poor clustering, where there is no relationship between the clusters and the true progenitor groups.

\item the Density-Based Clustering Validation index \citep[DBCV,][]{DBCV}, an internal metric that does not require information on the true progenitor groups. DBCV is specifically designed for density-based methods and evaluates the intrinsic quality of a clustering by assessing the density within clusters relative to the density between clusters. It rewards partitions where clusters are internally dense (low sparseness) and well separated from other clusters by low-density regions.
\end{enumerate}

By optimizing the \texttt{HDBSCAN} parameters using the V-measure, we establish a performance benchmark that represents the best possible clustering outcome with respect to the known progenitor labels. The DBCV metric is then used to investigate whether a comparable performance can be achieved in the more realistic scenario where the true origin of the accreted debris is unknown.

\subsection{Evaluation metrics}

While the V-measure provides a global assessment of the purity and completeness of clustering, we also investigate the purity and completeness for the individual clusters. To this end, we calculate the \textit{precision} and \textit{recall} for each identified cluster with respect to its dominant progenitor group (i.e., the progenitor with the most members in that cluster). Here, precision is defined as the fraction of the members of a cluster that belong to the dominant progenitor, while recall is the fraction of all stars from that dominant progenitor that are contained within the cluster. Assuming $\mathbf{Cl}$ is the set of particles of a cluster and $\mathbf{Pr}$ is the set of particles from the dominant progenitor of that cluster, i.e., sharing the same \texttt{PeakMassIndex}, the precision and recall scores for the cluster are calculated as:

\begin{equation*}
    \mathrm{Precision}= {P} = \frac{| \mathbf{Pr}\cap \mathbf{Cl}|}{|\mathbf{Cl}|} \;\;\;\; \text{and} \;\;\;\; \mathrm{Recall}={R}=\frac{| \mathbf{Pr}\cap \mathbf{Cl|}}{|\mathbf{Pr}|},
\end{equation*}
where ``$|\;\;|$'' indicates the cardinality, i.e., the number of elements of the sets.

In order to quantify the {\it spread} of the stellar debris of a progenitor in a multidimensional parameter space (see Section~\ref{sec:clustering_approach}), we calculate its covariance matrix $(\mathcal{C})$ (from now on we will refer to the covariance matrix as the ``spread''). This parameter captures the information on both the dispersion along the single properties of the stellar debris and their inter-dimensional correlations. The trace of the covariance matrix is essentially the \textit{total variance} $(\sigma_{T}^{2})$ and measures the spread of values for the particles in the distribution, i.e.,
 \begin{equation*}
    \sigma_{T}^{2}=\mathrm{tr}(\mathcal{C})=\Sigma_{n=1}^{D}\,\sigma_{n}^{2},
 \end{equation*}
where $\sigma_{n}^{2}$ is the variance of the star particles along the dimension $n$.

Under the assumption that the distribution of the particles from a progenitor is a multivariate Gaussian in the parameter space, which is enforced by the transformation of variables in the preprocessing stage (see Section~\ref{sec:clustering_approach}), the degree of similarity between the debris from two different progenitors ($A$ and $B$) can be estimated using the Mahalanobis distance, i.e.,
\begin{equation*}
    \mathrm{D_{M}} = \sqrt{(\mathbf{\mu_{A}}-\mathbf{\mu_{B}})^{T}(\mathcal{C}_{A}+\mathcal{C}_{B})^{-1}(\mathbf{\mu_{A}}-\mathbf{\mu_{B}})},
\end{equation*}
where $\mathbf{\mu_{A}}$ and $\mathbf{\mu_{B}}$ represent the mean vectors of the stellar properties for the particles from progenitors $A$ and $B$. The Mahalanobis distance measures the difference between the mean stellar properties of the debris from the two progenitors in terms of their combined variance. We also use this metric to quantify the mixing of the debris of a progenitor with the in situ population.

\subsection{The 12-dimensional parameter (feature) space}
\label{sec:clustering_approach}

We apply the \texttt{HDBSCAN} algorithm on a multidimensional parameter space comprised of ages, chemical abundance ratios and kinematics of halo stars.
The kinematics are computed in a Cartesian coordinate system centred on the centre-of-mass of the host galaxy, and the $z$-axis aligned with the total angular momentum vector of the disc component. Two of the kinematical parameters are the integrals of motion (IoM): the energy $E$ and $L_{z}$, the component of the angular momentum along the $z$-axis, which is perpendicular to the plane of the stellar disc. We also include the perpendicular component of the angular momentum, which is the component in the $x-y$ plane, $L_{\perp}=\sqrt{L_{x}^{2}+L_{y}^2}$. This is often used, even though it is known to not be fully conserved \citep{Helmi_1999b}. To these, we add the Cartesian positions and velocities\footnote{We experimented with the removal of positions and velocities from the clustering features, but we found a decrease in the performance of the model compared with including them. We therefore decided to incorporate positions and velocities among the features as they may add information (for example, the location of particles in the host galaxy, which otherwise might be hard to disentangle from the angular momentum features).} ($x,y,z,v_{x},v_{y},v_{z}$). For chemical abundances, we use the $\alpha$-to-iron and iron-to-hydrogen chemical abundance ratios ($[\alpha/\mathrm{Fe}]$, $[\mathrm{Fe/H}]$), while for ages we use the lookback time since formation of star particles ($\tau_{\rm form}$). In summary, we choose a 12 dimensional (12-D) parameter space comprised of:

$\{{E, L_z, L_{\perp}, x, y, z, v_x, v_y, v_z, {\rm [Fe/H]}, [\alpha/{\rm Fe}], \tau_{\rm form}} \}$

Each of these stellar properties provides complementary information in the identification of progenitors. The IoMs are conserved quantities over time, which makes them ideal for identifying early mergers that appear as phase-mixed substructures at $z=0$ \citep{Helmi_1999}. The phase-space coordinates are most effective for recently accreted debris or progenitors that are actively being disrupted, where the positions and velocities can be highly correlated \citep{johnston_2008}. Chemical abundance ratios are excellent tracers of the common chemical evolution of stars within satellite galaxies prior to accretion onto their host \citep{FreemanBlandHawthorn_2002}, while stellar ages allow the reconstruction of the star formation history from a single progenitor \citep{Montalban_2021, Gallart_cmdfitting_2024}.

Because the \texttt{Auriga} galaxies include a mix of stellar populations, from phase-mixed stars in the inner galaxy to dynamically cold stellar streams in the outer halo, the feature distributions are often skewed and heavy-tailed along the single dimensions in the 12-D clustering space. To mitigate this, we preprocess the data to produce more Gaussian-like distributions. Specifically, we apply a cubic-root transformation to $L_z$, a logarithmic transformation to $L_{\perp}$, and an inverse hyperbolic sine transformation to the Cartesian coordinates ($x, y, z$). Finally, to account for the different scales and units of the various features, the data set of each galaxy is standardized by subtracting the median from each feature and dividing by its interquartile range, making it less sensitive to outliers than the standard scaling.

After preprocessing the data, we determine the optimal configuration of the \texttt{HDBSCAN} model parameters for each galaxy by exploring $640$ different configurations of model parameters, as described in Section~\ref{sec:hdbscan_tuning}.

\section{Results of Clustering}
\label{sec:results}

In this section, we present the results of clustering for both the accreted-only and accreted + in situ stellar haloes. These are presented separately, in 
subsections~\ref{sec:accreted_only} and ~\ref{sec:with_in situ}, respectively. The results on accreted-only haloes allow us to evaluate to what degree the assembly history of a galaxy can be reconstructed by \texttt{HDBSCAN} in the ``best-case scenario", without contamination from in situ stars. For the accreted + in situ case, we aim to exclude the stellar disc which otherwise would dominate over other overdensities in feature space. For this, we employ a velocity cut that separates halo stars from a rotationally-supported disc \citep[e.g.,][]{Koppelman_2019}; specifically, we define halo stars those with $|\mathbf{V}-\mathbf{V_{*}}| \geq 200 \, \text{km s}^{-1}$, where $\mathbf{V_{*}}=(0,200,0) \, \text{km s}^{-1}$ in a galactocentric cylindrical coordinate system $(\hat{v}_r, \hat{v}_{\theta}, \hat{v}_z)$. We note that even with this selection, the resulting data sets have considerably high in situ fractions, of $0.49$, $0.77$ and $0.72$ for Au7, Au25 and Au27, respectively.

The structure of the two subsections is very similar; they both include a discussion of the purity and completeness of the retrieved clusters and a comparison between the merger trees inferred from the clustering analysis with the full merger trees constructed from the original simulations.

\subsection{Clustering in accreted-only stellar haloes}
\label{sec:accreted_only}

\begin{figure*}
    \centering
    \begin{subfigure}{0.7\textwidth}
        \centering
        \includegraphics[width=\textwidth]{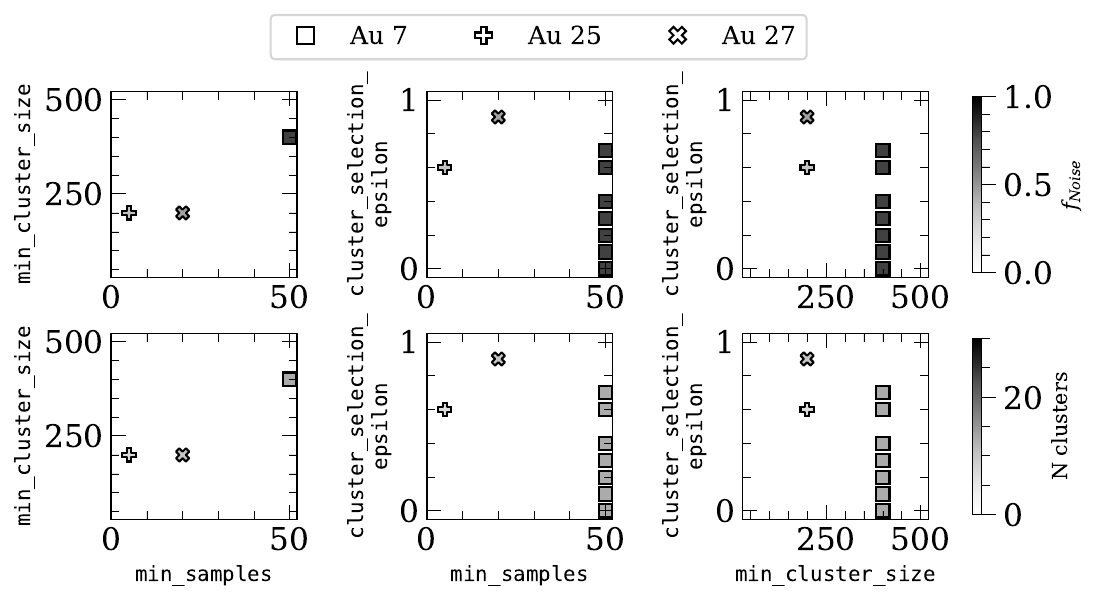}
        \caption{V-measure}
        \label{fig:best_params_vmeasure}
    \end{subfigure}
    \begin{subfigure}{0.7\textwidth}
        \centering
        \includegraphics[width=\textwidth]{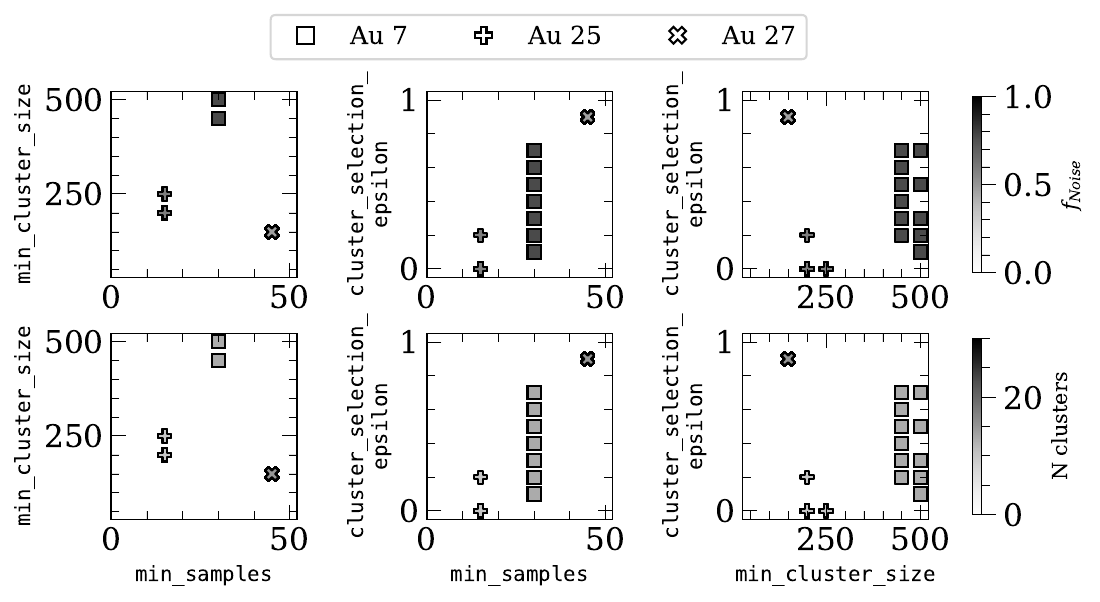}
        \caption{DBCV}
        \label{fig:best_params_DBCV}
    \end{subfigure}
    \caption{Optimal \texttt{HDBSCAN} parameter configurations found by the \texttt{Optuna} search. Panels (a) and (b) show the results of optimizations using the V-measure and DBCV metrics, respectively. In both panels, each symbol represents a different galaxy. The top sub-panels show the configurations coloured by the fraction of particles classified as noise, while the bottom sub-panels show the same configurations coloured by the number of clusters in the resulting partition.}
    \label{fig:optuna_results}
\end{figure*}

From the optimization procedure described in Section~\ref{sec:hdbscan_tuning}, we find that there are several ``optimal'' \texttt{HDBSCAN} parameter  configurations that return similar partitioning of the data. This can be seen by the consistent number of clusters and the fraction of noise points in Fig.~\ref{fig:optuna_results}, which shows the \texttt{HDBSCAN} parameter values for the runs performed with both the V-measure (\ref{fig:best_params_vmeasure}) and DBCV (\ref{fig:best_params_DBCV}) metrics. Configurations corresponding to a given host galaxy are shown with the same symbol. The number of clusters and the fraction of noise points associated with each clustering model are indicated on the greyscale bars.  This figure also shows that a degeneracy in the clustering evaluation metrics arises once the optimal values for the \texttt{min\_samples} and \texttt{min\_cluster\_size} parameters are identified. These two parameters establish an optimal condensed hierarchy, after which the final flat partition becomes less sensitive to the precise value of \texttt{cluster\_selection\_epsilon}, as the clusters can persist over a wide range of distance scales in the hierarchy. Based on these findings, we proceed to randomly select one configuration from this optimal set of parameter values to define the \texttt{HDBSCAN} model used for clustering.

\begin{figure*}
    \centering
    \includegraphics[width=0.8\textwidth]{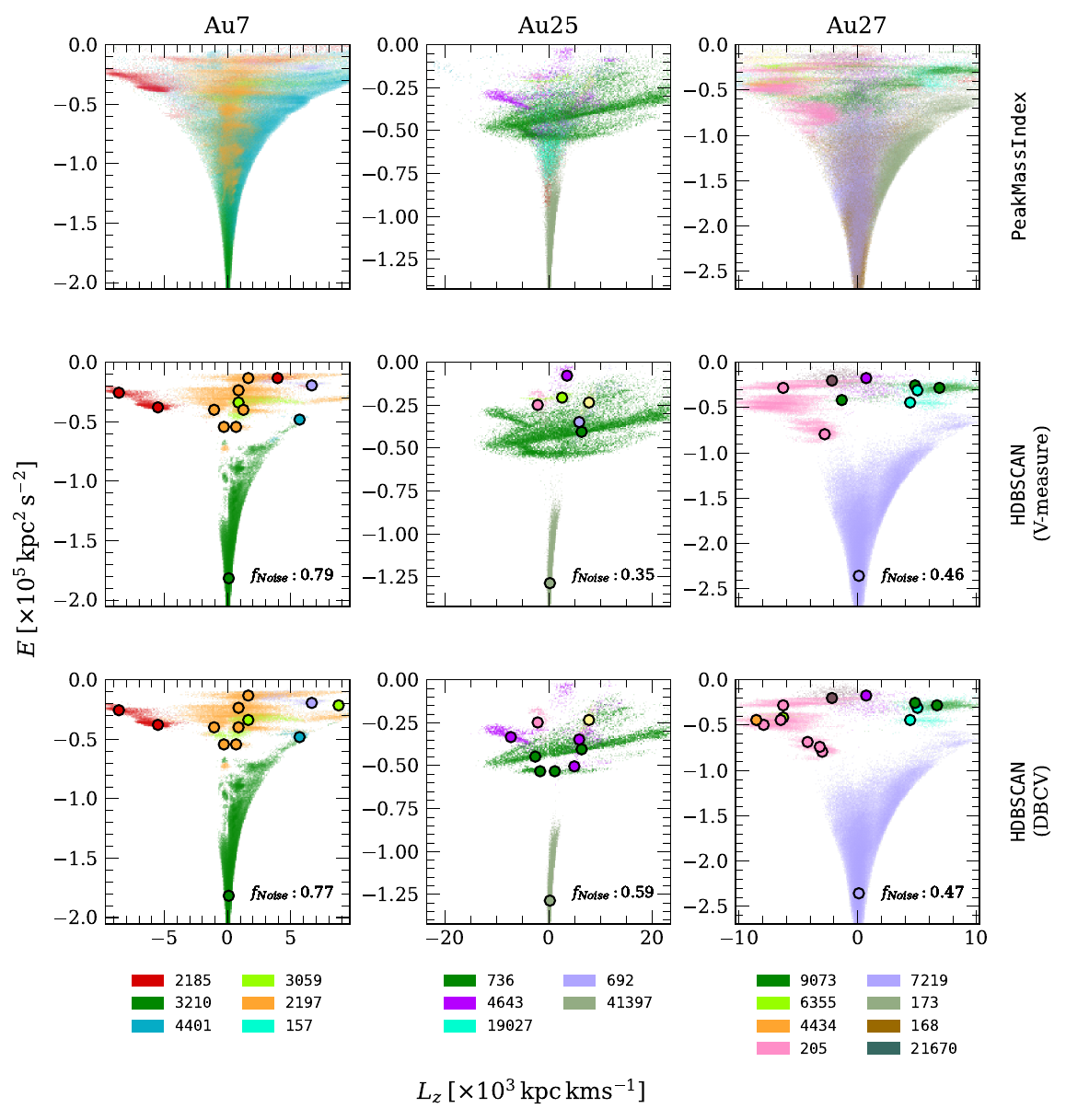}
    \caption{Density distribution of accreted star particles in the $E-L_z$ plane. The top row shows the ground-truth distribution, where particles are coloured by their progenitor group (\texttt{PeakMassIndex}). The middle and bottom rows show the \texttt{HDBSCAN} clustering results when optimized with the V-measure and DBCV metrics, respectively. In these lower panels, identified clusters are coloured according to their majority progenitor group, with noise particles shown in grey. The overplotted scatter points highlight the highest-density region of each cluster. The fraction of noise points is also included in each panel. At the bottom of each column, a legend shows the colours associated with the merger events reported in Table~\ref{tab:galaxy_sample}.}
    \label{fig:IoM_accreted}
\end{figure*}

The results of the clustering analysis are visualized in Fig.~\ref{fig:IoM_accreted}, which shows the density distribution of the accreted star particles in the $E-L_{z}$ plane for the three galaxies (one per column). The top row serves as the ground truth, with particles coloured according to their original progenitor group (i.e., their \texttt{PeakMassIndex}).  The middle and bottom rows show the results of the \texttt{HDBSCAN} clustering when optimized using the V-measure and DBCV metrics, respectively. To facilitate the comparison, each identified cluster is assigned the colour of the ground truth progenitor group that constitutes the majority of its members; particles classified as noise are shown in grey. For a better visualization of the distribution of clusters, the highest-density region of each cluster is marked with scatter circles. This helps locate the smaller clusters that may be obscured by the background noise in this plane.

From the top row, it is clear that only a small number of progenitors contribute the majority of accreted stars in each galaxy. As a consequence, most of the clusters identified by the \texttt{HDBSCAN} models are dominated by particles from these major merger events, some progenitors being resolved into multiple distinct clusters. Although a significant fraction of accreted stars are classified as noise, samples from the major merging events are consistently recovered. For instance, in Au7, which has the highest fraction of noise points, the identified clusters contain samples from 6 of the 10 progenitors, which, however, together contributed $\sim95\%$ of the total accreted particles.

The cluster distributions obtained with the V-measure and DBCV optimizations (middle and bottom rows of Fig.~\ref{fig:IoM_accreted}) are remarkably similar. This suggests that a partition of the data maximizing the intrinsic density and separation of the clusters is also strongly linked to one with a high level of purity and completeness. Quantitatively, the V-measure scores for the clusters obtained by optimizing with the V-measure are 0.17, 0.49, and 0.26 for Au7, Au25, and Au27, respectively. The partitions found by optimizing the DBCV index achieve comparable V-measure scores of 0.17, 0.37, and 0.24 for the same galaxies. This compares to V-measure scores of 0.01, 0.03 and 0.14 for the three galaxies when the clustering is performed with a non-optimized model (i.e., run with the default parameter values). This result is significant, as it suggests that even without ground-truth information, an unsupervised metric like the DBCV index can effectively guide the \texttt{HDBSCAN} algorithm toward a physically meaningful partition of the data that reflects the underlying progenitor structure. However, we note that the model optimized with the DBCV index sometimes tends to fragment progenitor groups into a larger number of clusters than the one optimized with the V-measure (see the green clusters for Au25 and the pink clusters for Au27 in Fig.~\ref{fig:IoM_accreted}).

\subsubsection{Purity and completeness of clusters}

\begin{figure}
    \centering
    \begin{subfigure}{\columnwidth}
        \centering
        \includegraphics[width=\textwidth]{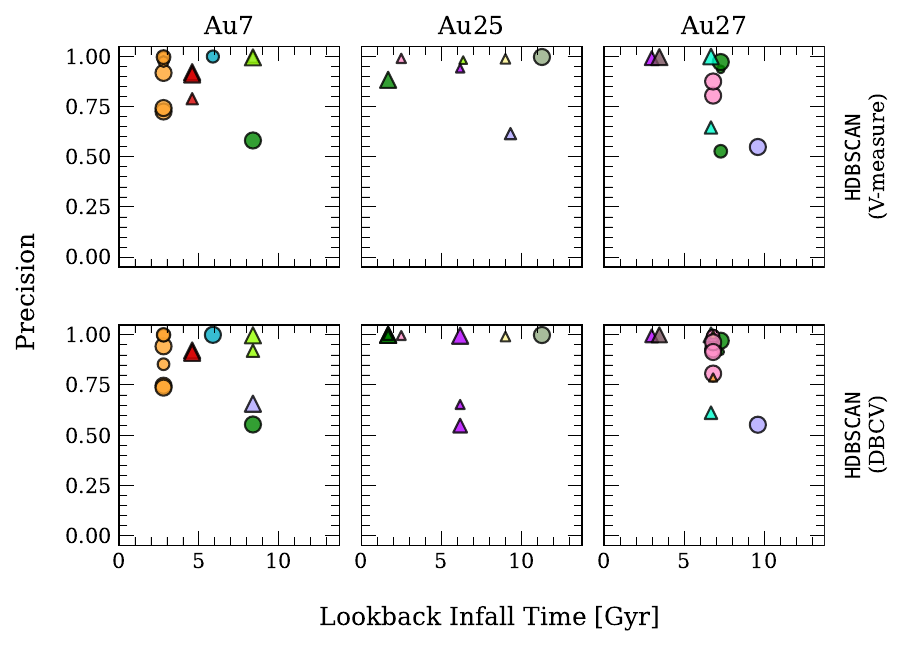}
    \end{subfigure}
    \begin{subfigure}{\columnwidth}
        \centering
        \includegraphics[width=\textwidth]{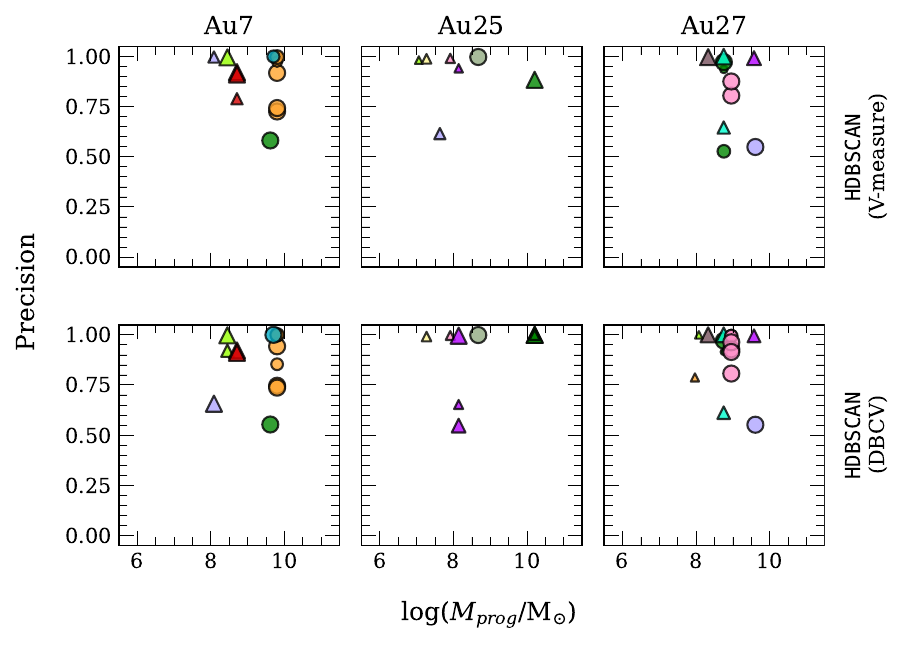}
    \end{subfigure}
    \caption{Precision scores for the clusters identified in accreted-only haloes by the \texttt{HDBSCAN} models optimized with the V-measure and DBCV index metrics. For each cluster, the precision is calculated with respect to the (dominant) progenitor with the highest number of cluster members, and plotted against the lookback infall time (top) and stellar mass at infall of the progenitor (bottom). The clusters are represented as triangles if the dominant progenitor is a stellar stream at $z=0$ and by circles if a phase-mixed structure. The size of the symbols is proportional to the number of members in the cluster, and the colour scheme follows the one in Fig.~\ref{fig:IoM_accreted} and indicates the dominant progenitor of each cluster.}
    \label{fig:P_accreted}
\end{figure}

\begin{figure}
    \centering
    \begin{subfigure}{\columnwidth}
        \centering
        \includegraphics[width=\textwidth]{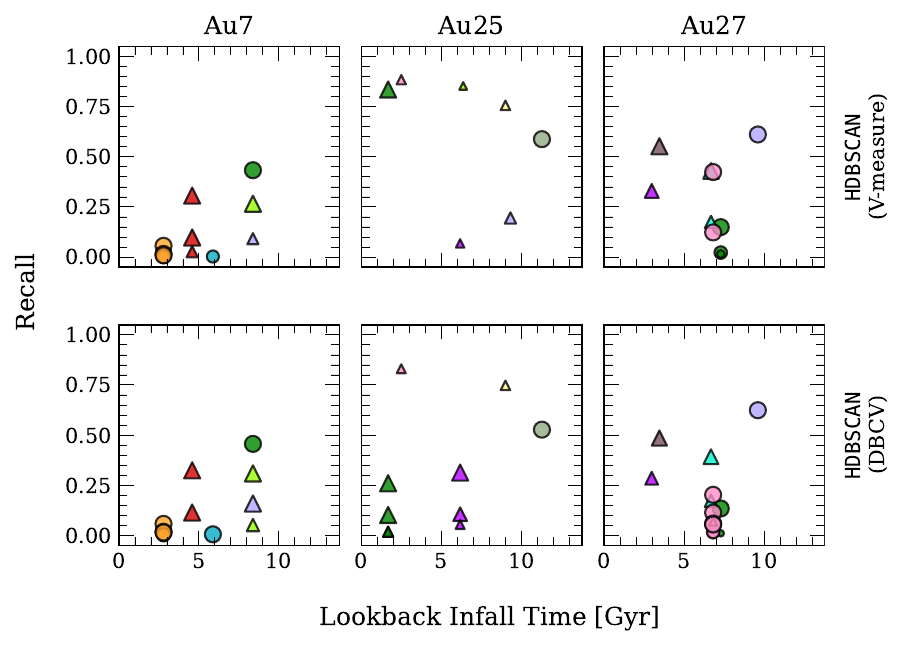}
    \end{subfigure}
    \begin{subfigure}{\columnwidth}
        \centering
        \includegraphics[width=\textwidth]{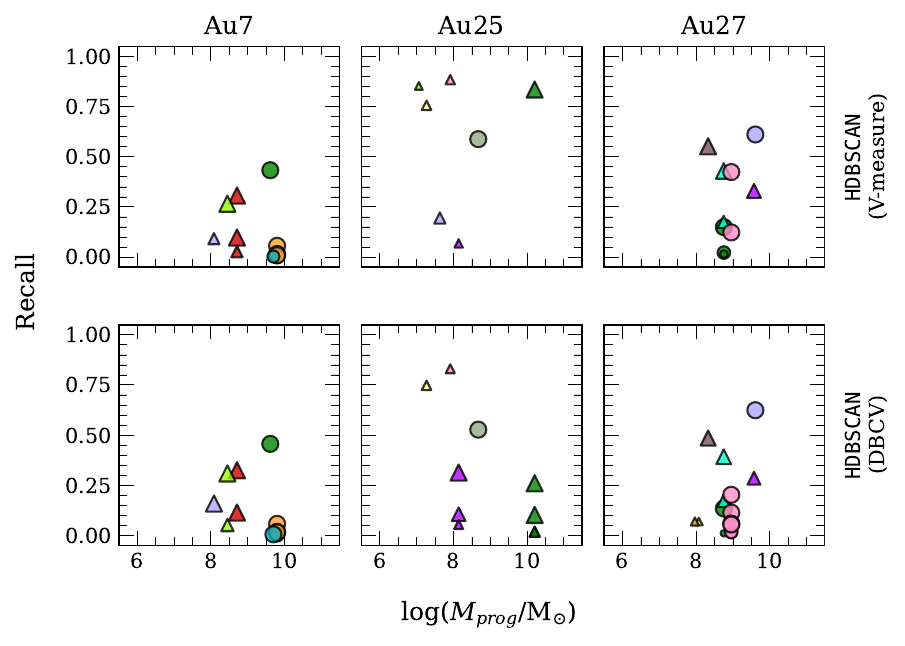}
    \end{subfigure}
    \caption{Recall scores for the clusters identified in accreted-only haloes by the \texttt{HDBSCAN} models optimized with the V-measure and DBCV index metrics. The symbols, colours and sizes are the same as described in Fig.~\ref{fig:P_accreted}, but now for the recall scores.}
    \label{fig:R_accreted}
\end{figure}

The precision and recall for the clusters identified in the sample galaxies are reported in Fig.~\ref{fig:P_accreted}  and Fig.~\ref{fig:R_accreted}, respectively. In each figure, the top panel refers to the clusters from the V-measure optimized model, while the bottom panel refers to the ones from the DBCV-optimized. The metrics are plotted against the lookback infall time, $\tau_{\rm infall}$ (top) and the stellar mass at infall (bottom) of the dominant progenitor. Each cluster is represented by a symbol that indicates whether its dominant progenitor is a stellar stream (triangles) or a phase-mixed substructure (circles) at $z=0$. The sizes of symbols scale with the number of stars in each cluster, the largest symbols representing clusters with more than 1000 members.

Generally, \texttt{HDBSCAN} identifies clusters with a high-precision, which means that clusters are mostly composed of stars that belong to a single \textit{dominant} progenitor. The lowest precision scores for the V-measure model are associated with specific physical scenarios. For Au7, the cluster linked to prog.~\texttt{3210} (precision $P=0.58$) corresponds to the oldest major merger. In Au25, the cluster for prog.~\texttt{692} ($P=0.62$) is a small stream dynamically similar to debris from a recent major merger, prog.~\texttt{736}. For Au27,low-precision scores are found for clusters associated with prog.~\texttt{9073} ($P=0.53$), prog~\texttt{248} ($P=0.65$) and prog.~\texttt{7219} ($P=0.55$). The former two progenitors are fragmented into three and two clusters, respectively; the low-precision clusters are also those that overlap with other debris in the phase space. The cluster associated with prog.~\texttt{7219} is significantly contaminated by debris from the other two main merger events (prog.~\texttt{168} and prog.~\texttt{173}), which slightly preceded the dominant progenitor and have not been recovered by the clustering model. The DBCV-optimized model identifies the same low-precision progenitors, with the addition of clusters for prog.~\texttt{4643} in Au25, which also dynamically overlaps with debris from prog.~\texttt{736}.

On average, clusters associated with stellar streams show higher precision ($P=0.91$ for V-measure and $0.90$ for DBCV) than those associated with phase-mixed structures ($P=0.84$ for V-measure and $0.88$ for DBCV). These dynamically cold filamentary structures are less likely to overlap with debris from other progenitors than the phase-mixed structures, the latter often being contaminated by earlier mergers that remain undetected. This effect may also explain the higher precision found for clusters from the late mergers (that is, lookback time of infall $\leq  5$~Gyr) compared to earlier ones; this effect is particularly evident for Au27 (see the rightmost subpanel in top panel of Fig.~\ref{fig:P_accreted}). In terms of progenitor masses, clusters from the most massive mergers tend to have lower purity scores, which is likely due to the wider range in stellar properties (features) of their debris.

In contrast to the high-precision scores, the recall scores are significantly lower for both models. The average recall ($R$) values for clusters of stellar streams are 0.40 (for V-measure) and 0.24 (for DBCV), while for clusters of phase-mixed structures are 0.16 (for V-measure) and 0.13 (for DBCV), respectively. This is primarily due to two factors: a large number of particles are being classified as noise, and the fragmentation of single progenitors into multiple clusters. The latter effect explains the lower recall of the DBCV-optimized model, which tends to produce more clusters per progenitor. For example, for prog.~\texttt{2197} in Au7, $85\%$ of its particles are classified as noise in both models, while $10\%$ are split into six separate clusters.

\begin{figure}
    \centering
    \includegraphics[width=0.8\linewidth]{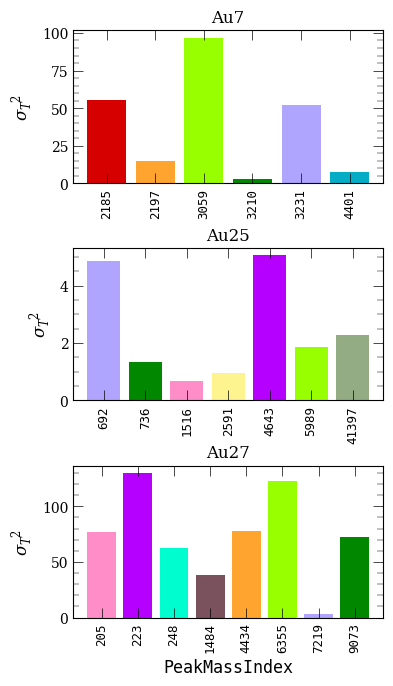}
    \caption{Total variance values for the progenitor galaxies appearing as dominant progenitors in the \texttt{HDBSCAN} clusters.}
    \label{fig:total_variance}
\end{figure}

\begin{figure}
    \centering
    \includegraphics[width=0.8\linewidth]{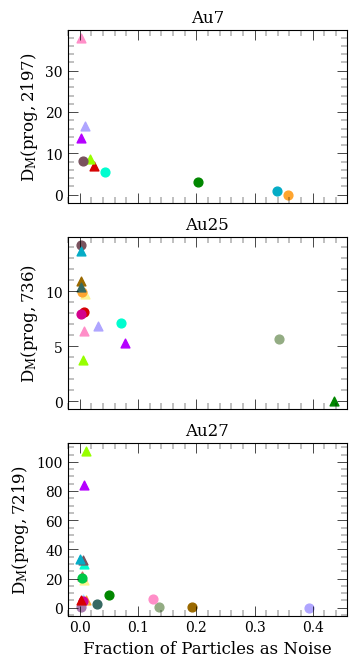}
    \caption{Mahalanobis distance between the 12-D debris distributions of the progenitor with the largest number of particles classified as noise and the other progenitors of the galaxies. The symbol and colours of the progenitors are the same one used in Fig.~\ref{fig:P_accreted}. }
    \label{fig:DM_noise}
\end{figure}

The fragmentation of the debris of a progenitor into multiple clusters is expected for the most massive progenitors.  Those are typically accreted late and experience a prolonged chemical evolution; therefore, they tend to consist of multiple stellar populations and their debris spans a wide range in chrono-chemodynamical properties. To investigate this hypothesis, we calculated the total variance of the debris of each progenitor in the 12-dimensional parameter space (Fig.~\ref{fig:total_variance}). Progenitors that are split into several clusters generally show higher variance than the rest. However, a large variance does not always lead to fragmentation. In cases where the debris are too widely dispersed, the algorithm either assigns many of those stars to other clusters or classifies them as noise.
Stars are most often classified as noise when debris from different progenitors overlap in the parameter space. This effect is illustrated in Fig.~\ref{fig:DM_noise}, where the progenitors with the largest number of stars flagged as noise are also those with the smallest Mahalanobis distance from the progenitor that dominates the noise population.

\subsubsection{The assembly histories inferred from clustering}

\begin{figure*}
    \centering
    \includegraphics[width=0.7\textwidth]{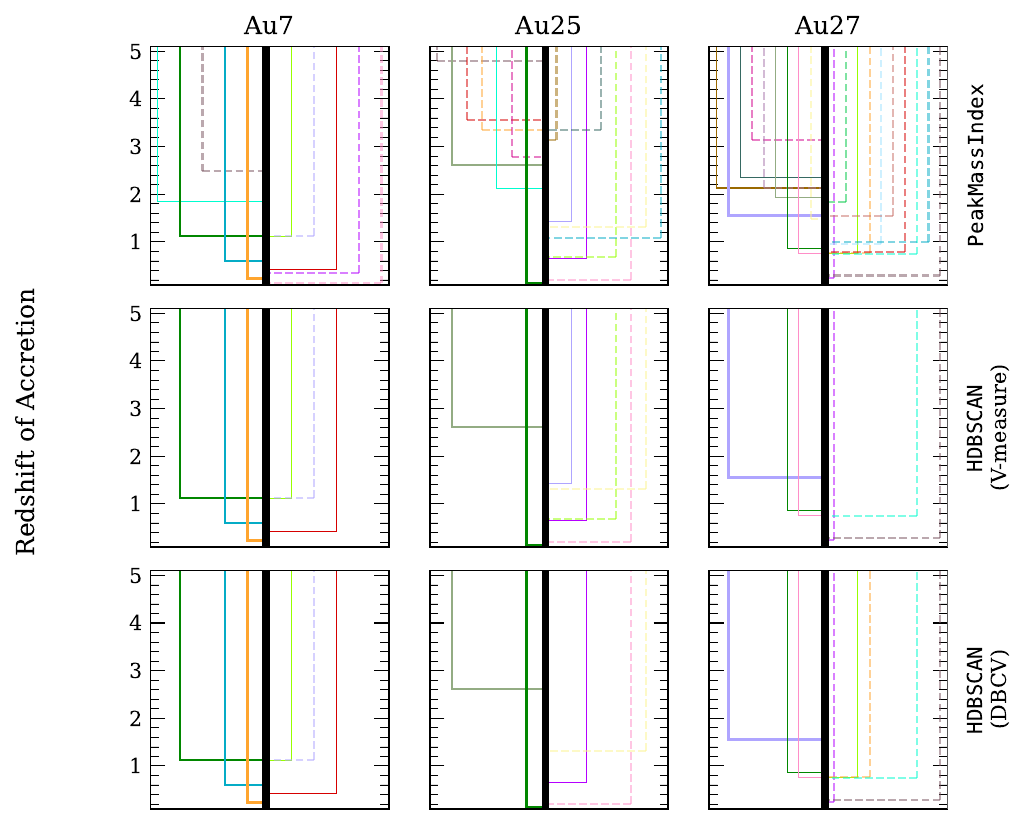}
    \caption{Merger trees for the galaxies in the sample as reconstructed by the V-measure (middle) and DBCV (bottom) optimized \texttt{HDBSCAN} models for the accreted-only haloes. Top panel shows the full merger trees. Each colour represents a different progenitor galaxy. The thickness of each line is proportional to the number of accreted star particles from that progenitor. Dashed lines indicate progenitors that contribute $<1\%$ to the overall accreted population of the galaxy at $z=0$.}
    \label{fig:merger_tree_acc}
\end{figure*}

We further investigate the assembly histories of the three galaxies, as represented by their merger trees constructed from the clusters recovered by the optimized algorithm. The full merger trees for the sample galaxies, computed directly from the simulations with the methods described in Section~\ref{sec:data}, are shown in the top panel of Fig.~\ref{fig:merger_tree_acc}. Each branch in the tree represents a progenitor galaxy that joins the main branch (shown in black) when the progenitor falls into the host. The thickness of each branch is proportional to the number of accreted particles that the progenitor satellite contributes to the host at $z=0$. The dashed lines indicate progenitors that contribute less than 1\% of the total number of accreted particles at $z=0$. The middle and bottom panels show the merger trees as inferred from the clustering results obtained with the V-measure-optimized and DBCV-optimized \texttt{HDBSCAN} models, respectively. Here, a branch is included only if its corresponding progenitor is dominant in at least one identified cluster.

As expected, both models are more successful in recovering more massive mergers (those contributing >1\% of all accreted particles, shown as solid lines and reported in Table~\ref{tab:galaxy_sample}).  The major merger events that are missed are predominantly those that occur at high redshift ($z_{\rm acc} \gtrsim 2$), with the exception of prog.~\texttt{6355} (light green) in Au27. This progenitor merged with the main galaxy around the same time as prog.~\texttt{205}, a satellite about ten times as massive; thus, its stellar debris was likely masked in the feature space by the dominant signature of the larger satellite, preventing its identification by the clustering algorithm.


\subsection{Clustering in accreted + in situ stellar haloes}
\label{sec:with_in situ}


\begin{figure*}
    \centering
    \includegraphics[width=0.8\textwidth]{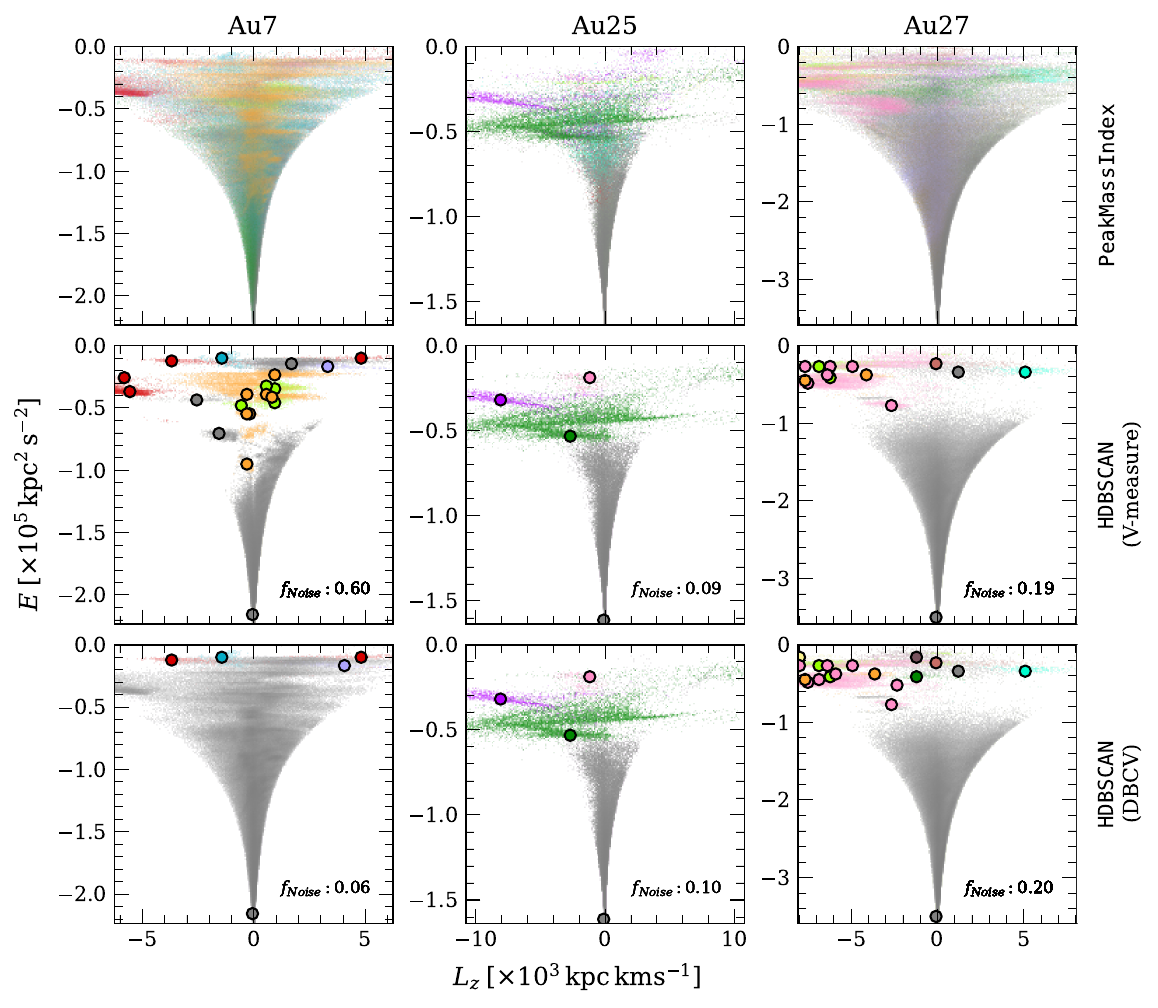}
    \caption{Density distributions in the $E-L_z$ plane for the accreted stars (similar to Fig.~\ref{fig:IoM_accreted}), but now including the in situ stars which are shown in grey. The clustering results from the optimized models are shown in the middle panel for the V-measure and the bottom panel for DBCV. The noise distributions are not shown, however we report the fraction of points classified as noise in each panel.}
    \label{fig:IoM_insitu}
\end{figure*}

Fig.~\ref{fig:IoM_insitu} shows the distribution of the \texttt{HDBSCAN} clusters in the $E-L_z$ plane obtained from the data set contaminated with in situ stars, which is a similar analysis done in Fig.~\ref{fig:IoM_accreted} for the accreted-only case. In this figure, however, the noise points are not shown to avoid confusion with the in situ stars which are shown in grey. However, we include the fraction of points classified as noise by both V-measure and DBCV models. The figure also adopts the same colour scheme as used previously in Fig.~\ref{fig:IoM_accreted}.

Several observations can be made from the results of Fig.~\ref{fig:IoM_insitu}. First, the fraction of points classified as noise decreases substantially in comparison with the accreted-only case, as now the dense in situ population forms its own clusters. At the same time, the number of recovered accreted progenitors also decreases, as the particles belonging to these missing progenitors end up in the in situ clusters. Despite this contamination, the majority of identified clusters are still dominated by stars from a single accreted progenitor, and, as in the accreted-only analysis, the partitions from the V-measure and DBCV optimizations are generally very similar. A notable exception is Au7, where the V-measure-optimized model yields a significantly different partition (20 clusters, $f_{\text{noise}}=0.60$) compared to the DBCV-optimized model (5 clusters, $f_{\text{noise}}=0.06$). The key difference in the optimal parameters for Au7 lies in the \texttt{min\_samples} value: 10 for the V-measure-optimized model and 5 for the DBCV-optimized one. By increasing \texttt{min\_samples}, the density estimate of the algorithm becomes more conservative, requiring more neighbours for a point to be considered part of a dense core. The different clustering configurations resulting from the two values of \texttt{min\_samples} can be explained in terms of the overlap in the clustering space between the oldest debris mergers and the in situ population. 

\begin{figure}
    \centering
    \includegraphics[width=0.8\linewidth]{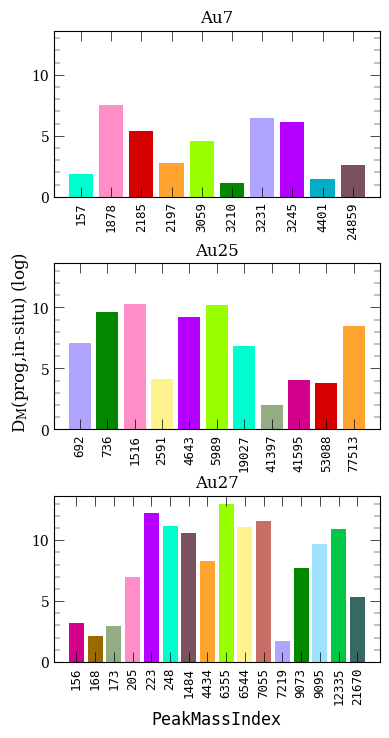}
    \caption{Mahalanobis distance between the particle distributions of the progenitor galaxies and the in situ population. Colours are associated to progenitors using the same colour scheme as in Fig.~\ref{fig:IoM_accreted}. }
    \label{fig:DM_insitu}
\end{figure}

Fig.~\ref{fig:DM_insitu} shows the Mahalanobis distance between the 12-D distributions of the debris from the progenitors and the in situ population. This shows that the oldest major mergers (e.g., prog.~\texttt{3210} in Au7, prog.~\texttt{41397} in Au25 and prog.~\texttt{7219} and in Au27) have the largest overlap with the in situ population in the clustering space. Therefore, the particles from these mergers tend to be associated with the same cluster of the in situ population. A stricter density requirement (i.e., a higher value of \texttt{min\_samples}) causes a fragmentation of the sparser regions of the in situ cluster. Because a larger number of progenitors in Au7 are mixed with the in situ ones than in the other galaxies, the fragmentation of the in situ clusters leads to the formation of groups dominated by accreted particles, hence increasing the overall V-measure score of the clustering configuration despite of a decrease in the cluster density.

\subsubsection{Purity and completeness of clusters}

\begin{figure}
    \centering
    \includegraphics[width=\columnwidth]{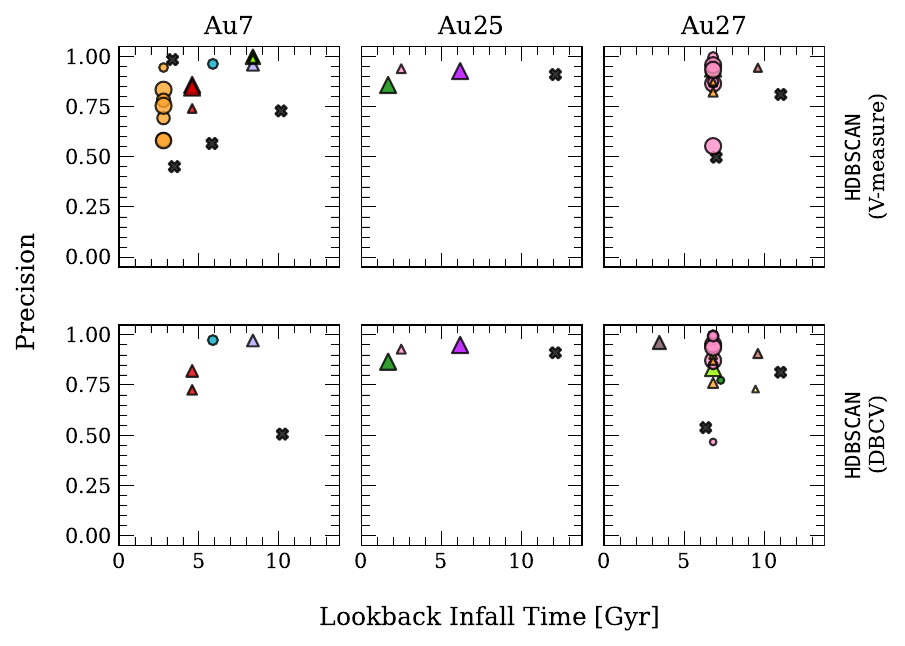}
    \caption{Precision scores for the clusters identified by the \texttt{HDBSCAN} models optimized with the V-measure and DBCV index metrics for the accreted + in situ haloes. Symbols and colours are the same as in Fig.~\ref{fig:P_accreted}, for the accreted-only haloes. Crosses indicate clusters that are dominated by in situ particles and their associated ``lookback infall time'' is the $90^{\rm th}$ percentile of the age distribution of the cluster.}
    \label{fig:P_insitu}
\end{figure}

\begin{figure}
    \centering
    \includegraphics[width=\columnwidth]{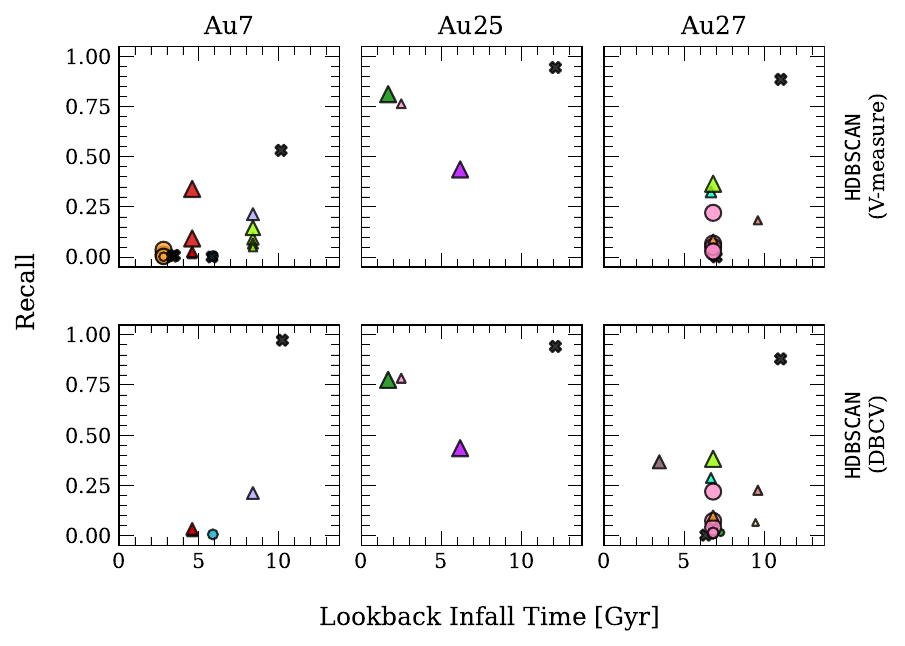}
    \caption{Recall scores for the clusters identified in accreted + in situ haloes by the \texttt{HDBSCAN} models optimized with the V-measure and DBCV index metrics. Symbols, sizes and colours are the same as in Fig.~\ref{fig:P_insitu}.}
    \label{fig:R_insitu}
\end{figure}

The purity and completeness of the clusters identified in the in situ contaminated data sets are described through the precision and recall scores in Figs.~\ref{fig:P_insitu} and \ref{fig:R_insitu}. Similarly to Figs.~\ref{fig:P_accreted} and \ref{fig:R_accreted}, the clusters are represented by triangles if the debris from the corresponding dominant progenitor is in a form of stellar stream at $z=0$, or by circles if it is in a phase-mixed structure. The colour of the cluster symbol is the same as used for the dominant progenitor in Fig.~\ref{fig:IoM_insitu}, while the size indicates the number of members in the cluster. We also show the clusters dominated by in situ particles as black crosses. In Figs.~\ref{fig:P_insitu} and \ref{fig:R_insitu}, the precision and recall scores are plotted against the lookback infall time of their dominant progenitors; for in situ clusters, this is taken as the $90^{\rm th}$ percentile of the age distribution of the particles in the cluster. If an in situ cluster were mistakenly to be considered as a sample of accreted stars, this statistic could be used as a proxy to the infall time of the relative progenitor galaxy.

As shown in Fig.~\ref{fig:P_insitu}, the contamination from in situ particles only slightly degrades the precision of the identified accreted clusters, with most still being composed of well over 50\% of stars from their dominant progenitor. The average precision for clusters associated with stellar streams and phase-mixed substructures is 0.91 and 0.81, respectively, for the V-measure optimized model, and 0.87 and 0.83 for the DBCV one. Conversely, the recall scores decreased on average: 0.23 (V-measure) and 0.27 (DBCV) for streams, 0.03 (V-measure) and 0.04 (DBCV) for phase-mixed structures. This is due to increased fragmentation, as shown in Fig.~\ref{fig:R_insitu}. A notable exception is Au25, whose larger progenitors, which appear as stellar streams, are recovered with similar or even higher recall scores than in the accreted-only analysis. This counterintuitive result likely arises from a contrast effect: the introduction of the dense, clustered in situ population makes the relatively less dense but coherent stellar streams stand out more clearly, allowing the algorithm to capture a larger fraction of their members. The in situ dominated clusters are also particularly pure and have high recall scores. This is probably due to both the efficiency of the algorithm of isolating the dense in situ component into one or a few coherent groups, and the significantly larger number of in situ particles compared to the accreted ones in these regions of the clustering space. The ``inferred'' lookback infall time for the in situ clusters, i.e., the $90^{\rm th}$ percentiles of the cluster age distributions, appears to vary significantly, which indicates that particles that formed in different star formation episodes also have different chrono-chemodynamical properties. In all galaxies, the only clusters with an associated lookback infall time $>10$~Gyr are in situ dominated.

\subsubsection{The assembly histories inferred from clustering}
\label{sec:assembly_history}

\begin{figure*}
    \centering
    \includegraphics[width=0.7\textwidth]{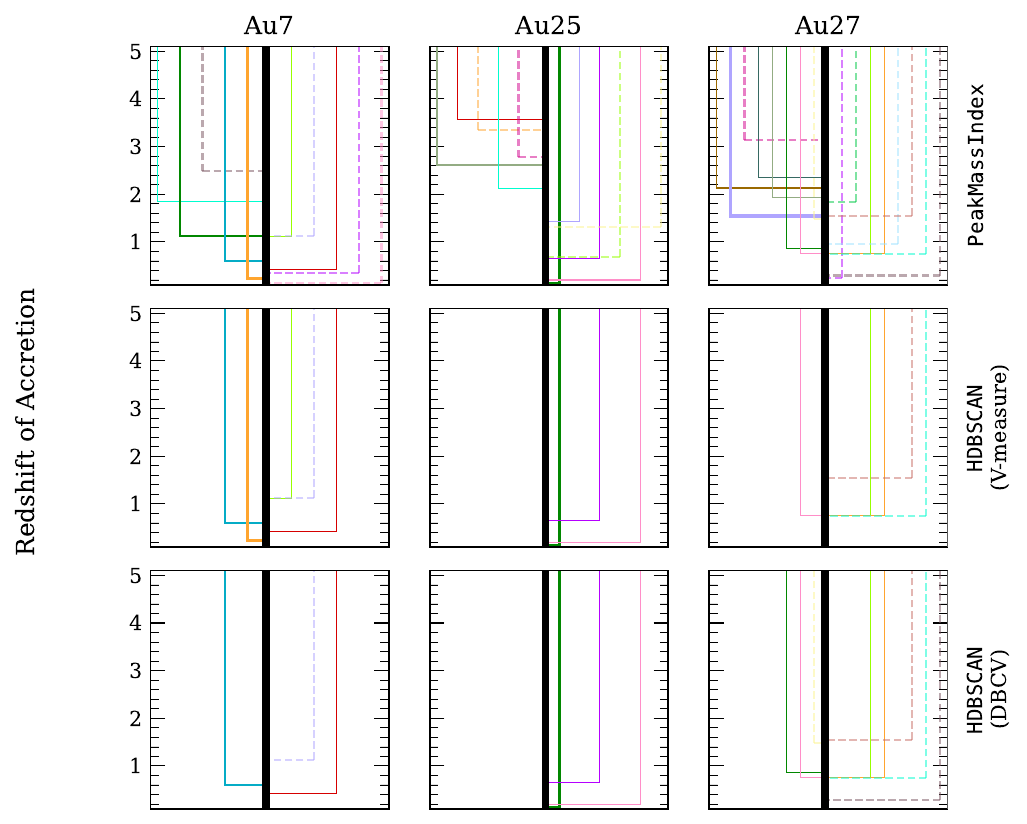}
    \caption{Merger trees for the galaxies in the sample as reconstructed by the V-measure (middle) and DBCV (bottom) optimized \texttt{HDBSCAN} models when contamination from in situ stars is included. Top panel shows the full merger trees. The correspondence between lines and progenitors is the same as described in Fig.~\ref{fig:merger_tree_acc} for the accreted-only case. The style and thickness of the lines reflects the new fraction of accreted stars from each progenitor after the velocity selection cut is applied.}
    \label{fig:merger_tree_insitu}
\end{figure*}

A comparison of the results with and without in situ contamination reveals that the primary impact of the in situ particles is not a degradation in the purity of recovered clusters, but rather the complete non-detection of certain progenitor groups. To better understand which mergers are more likely to be affected, we analyze the merger tree plots in Fig.~\ref{fig:merger_tree_insitu}. Similarly to Fig.~\ref{fig:merger_tree_acc}, the actual merger history of the galaxies is shown in the top panel, while the middle and bottom panels only display those merger events which have been identified as the dominant progenitors of the \texttt{HDBSCAN} clusters. Again, the continuous lines indicate merger events that contribute at least 1\% of the total accreted stars in the galaxy (after the velocity selection cut) at $z=0$, and the thickness of the line is proportional to the actual number. 

This figure shows that none of the major progenitors accreted before $z_{\rm acc}\sim1$ (i.e., $\sim8~\mathrm{Gyr}$ ago) are retrieved as clusters by the optimized \texttt{HDBSCAN} model. This suggests that the clustering model may identify only the cold stellar streams, as also seen in Figs.~\ref{fig:P_accreted} and \ref{fig:R_accreted}. Therefore, in the presence of in situ stars, the model can retrieve only the most recent fraction of the overall merger history of the galaxy. This result is in clear contrast to the results obtained in Section~\ref{sec:accreted_only} for accreted-only haloes, where the model could retrieve some of the most ancient merger events. Hence, reducing the contamination of in situ stars in the sample of halo stars is crucial for an effective recovery of old ($z_{\rm acc}>1$) merger events when using \texttt{HDBSCAN}.

\subsection{The impact of stellar ages}
\label{sec:no_ages}

As already mentioned, in our analysis clusters are obtained from the full chrono-chemodynamical properties of the accreted particles. However, replicating this parameter space for stars in the Milky Way might be challenging. Although kinematics measurements are accurately provided for tens of millions of stars by Gaia and self-consistent chemical abundances are becoming available for a growing number of stars through the combination of data from different spectroscopic surveys \citep{Thomas2024,horta_2025}, stellar ages remain accessible only for a fraction of them. Therefore, a more realistic application of our methodology would involve only the dynamical and chemical properties of the stars in the Milky Way. In order to assess the quality of the merger tree reconstruction under these conditions, we repeat the analysis discussed in Sections~\ref{sec:accreted_only} and \ref{sec:with_in situ} excluding the ages of the star particles ($\tau_{\text{form}}$) from the vector of stellar properties describing the clustering space. For the sake of conciseness, we consider only Au27, which is the galaxy with the assembly history most similar to the Milky Way in our sample, for this part of the analysis. 

\begin{figure}
    \centering
    \includegraphics[width=\columnwidth]{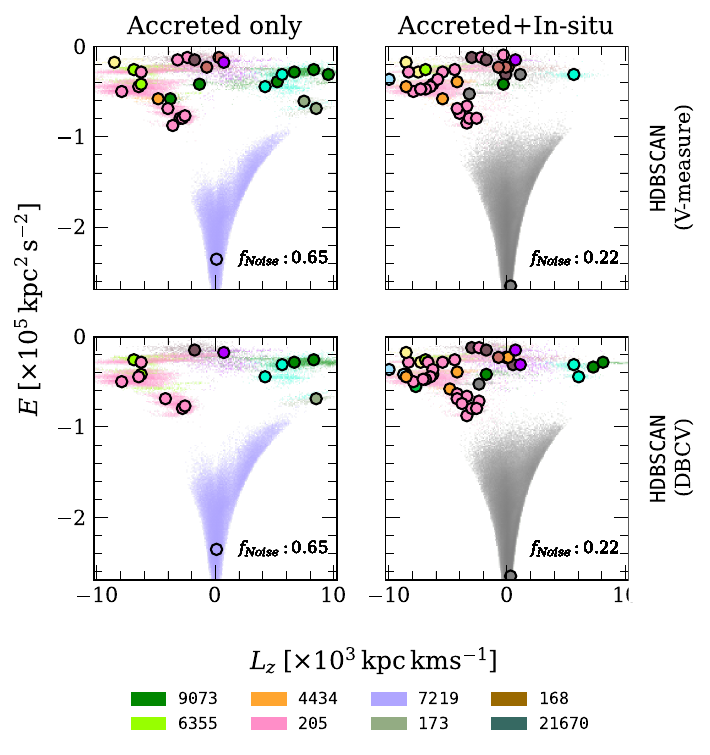}
    \caption{Density distribution in the $E-L_z$ plane for the stars in Au27 colour-coded by their \texttt{HDBSCAN} label. The clustering analysis is repeated excluding the age of the star particles from both the accreted-only and accreted + in-situ datasets. The plot format follows Figs.~\ref{fig:IoM_accreted} and \ref{fig:IoM_insitu}.  }
    \label{fig:IoM_noages}
\end{figure}

Similarly to Figs.~\ref{fig:IoM_accreted} and \ref{fig:IoM_insitu}, the distribution in the $E-L_z$ plane of the star particles in Au27 coloured based on their \texttt{HDBSCAN} group are reported in Fig.~\ref{fig:IoM_noages} for both the clustering considering only accreted stars (left plots) and including in-situ contamination (right plots). In both cases, the \texttt{HDBSCAN} parameters have been fine-tuned to adapt to the new clustering space without the age information. The top plots show the results of the clustering model optimised using the V-measure metric, while the bottom ones refer to the optimisation obtained with the DBCV metric. By comparing Fig.~\ref{fig:IoM_noages} with Figs.~\ref{fig:IoM_accreted} and \ref{fig:IoM_insitu}, it can be noticed that excluding the stellar ages results in a larger number of clusters. The clustering model optimised with V-measure returns 30 and 38 clusters in this scenario for the accreted and accreted~+~in-situ datasets, while only 11 and 15 clusters, respectively, are identified by the same model when the age information is provided. Similarly, the DBCV-optimised model returns a larger number of clusters for both the accreted-only (17 vs 16) and accreted~+~in-situ (46 vs 20) cases. This effect can be explained as fragmentation of the progenitors identified in the analysis with stellar ages into a larger number of clusters and a consequent decrease in the average completeness of the retrieved groups in both the accreted-only  (0.15 vs 0.29, V-measure and 0.24 vs 0.19, DBCV) and accreted~+~in-situ (0.08 vs 0.16, V-measure and 0.09 vs 0.15, DBCV) datasets. The average purity of the groups is similar whether the clustering is performed with (0.78, V-measure, and 0.86, DBCV, for the accreted-only stars and 0.88, V-measure, and 0.87, DBCV, for the accreted~+~in-situ stars) or without the age (0.90, V-measure, and 0.84, DBCV, for the accreted-only stars and 0.87, V-measure, and 0.87, DBCV, for the accreted~+~in-situ stars) of the star particles. Therefore, the identification of pure samples of stars accreted from a single progenitors is still possible without the knowledge of their formation time. However, the inclusion of such information is important for grouping the accreted stars in a number of substructures more representative to the number of actual progenitor galaxies. Therefore, if stellar ages cannot be determined, the development of a methodology for further grouping the star samples identified by \texttt{HDBSCAN} is crucial for a correct inference of the Galaxy assembly history.

\section{Discussion}
\label{sec:discussions}

Our analysis shows that an optimized \texttt{HDBSCAN} model can produce pure clusters of accreted stars even in environments strongly contaminated by the in situ population (see Fig.~\ref{fig:P_insitu}). This is in contrast with the results of \citet{Thomas2025} who use the single-linkage clustering methodology developed by \citet{Lovdal_2022} and find that it predominantly returns clusters composed in majority by in situ stars or by a combination of debris from different progenitors (see Section 4.1 in \citealt{Thomas2025}). We believe that the optimization of the clustering algorithm is the crucial factor in correctly identifying the clusters made of accreted stars. Also, while the \citet{Lovdal_2022} clustering algorithm uses only properties related to IoM, our study uses an extended, 12-D parameter (feature) space that combines chemistry, kinematics and ages. There are, of course, other factors that may explain the discrepancies between the results of the two methods. The \citealt{Lovdal_2022} methodology 
imposes the creation of ``background haloes'' in order to determine the significance score of the clusters, which may introduce artificial density variations in the clustering space. There are also differences between the two \texttt{Auriga} data sets in terms of their numerical resolution, as \citet{Thomas2025} use the ``level 3'' suite, which are higher in resolution than the ``level 4'' suite used in our analysis. As the stellar masses of galaxies, both in situ and accreted, increase at higher resolution (see e.g., \citealt{Grand_2021} and \citealt{riley_2025}), it is possible that the contamination by in situ stars in the sample also increases. Moreover, although both studies use samples of star particles with similar halo-type kinematics, \citet{Thomas2025} focus on solar neighbourhood regions, whereas we study the entire haloes. 

Our optimized \texttt{HDBSCAN} model can retrieve mergers up to $z_{\rm acc}\sim3$ (see Fig.~\ref{fig:merger_tree_acc}). However, when applied to a data set contaminated by in situ stars, only galaxies accreted after $z_{\rm acc} \simeq 1$ are retrieved as dominant progenitors of the clusters. From an observational point of view, this suggests that accretion events >10 Gyr old may not be found by clustering methods, or that any clusters from this type of events are likely dominated by in situ material. This suggests that observational samples representing early accretions in the Galaxy, such as  GES \citep{Helmi2018,Belokurov2018}, Kraken \citep{Kruijssen2019}, or Shiva \& Shakti \citep{Malhan2024} could be significantly contaminated by in situ stars.

These findings are similar to the results of \citet{Thomas2025}, who show that their methodology is effective for mergers that occurred in the last $6-7$~Gyr. However, because the purity of the \texttt{HDBSCAN} clusters does not change significantly between the accreted-only and accreted + in situ data sets, we argue that the presence of the in situ population in the clustering data set mainly impacts the number and type of detectable mergers. We note that the problem could be attenuated when using observations, as the \texttt{Auriga} simulated haloes may have a higher fraction of in situ stars than measured in the Milky Way \citep{Monachesi_auriga_2019}. Therefore, we expect that a higher fraction of the assembly history of the Milky Way could be retrieved with our methodology than implied by the results from clustering on accreted + in situ stars in Section~\ref{sec:with_in situ}. Moreover, there are other machine learning methods that can very efficiently separate in situ stars from accreted stars \citep{Ostdiek2020,Sante2024}, before applying any clustering methods. Using these methods in combination, one would expect a higher rate of success in finding substructures, even in the case of heavy contamination by in situ stars.

In Sections \ref{sec:accreted_only}--\ref{sec:with_in situ}, the clusters were obtained from the full (12-D) chrono-chemodynamical properties of the accreted particles. However, replicating this 12-D parameter space for stars in the Milky Way may be challenging. Although the kinematical measurements are accurately provided for tens of millions of stars by Gaia and self-consistent chemical abundances are becoming available for a growing number of stars through the combination of data from different spectroscopic surveys \citep{Thomas2024,horta_2025}, stellar ages remain accessible only for a fraction of them. In Section \ref{sec:no_ages} we examined a more realistic scenario where the ages were not available in the set of features. We found that, on average, the purity of the clusters was not affected, but it was a slight decrease in their completeness. Overall, the cluster became ever more fragmented. 

Furthermore, the methodology presented in this paper does not account for any uncertainty or incompleteness in the stellar properties. Hence, more studies are needed to understand the implications of this missing information on the capability of \texttt{HDBSCAN} to recover past mergers.

\section{Conclusions}
\label{sec:concl}

We have developed and validated a new methodology for optimizing the parameters of the \texttt{HDBSCAN} clustering algorithm for identifying tidal debris in Milky Way-mass galaxies. Our approach is designed to maximize the recovery of coherent structures in chrono-chemodynamical space, and we demonstrate that this optimization simultaneously enhances both the purity and completeness of the resulting stellar groups (see Figs.~\ref{fig:IoM_accreted} and \ref{fig:IoM_insitu}), ensuring they correspond to the debris of single progenitors.

To test the power of this method, we applied it to three simulated Milky Way-mass galaxies from the \texttt{Auriga} suite, which feature diverse assembly histories. We first established a baseline performance in an idealized scenario, using only accreted star particles. In these ideal conditions, we found that

\begin{itemize}
    \item The vast majority of clusters identified by \texttt{HDBSCAN} are dominated by stars originating from a single accreted progenitor, confirming the ability of the algorithm to trace individual merger events.
    \item Cluster members represent only a small fraction of the total number of the accreted particles associated to a progenitor, as particles are lost as noise points or, for massive systems, split into different clusters due to the wide range in the chrono-chemodynamical properties of their debris.
    \item The major mergers of a galaxy can be retrieved as clusters up to a redshift of accretion of $z_{\rm acc}\sim3$ unless a more recent massive merger has occurred (see Fig.~\ref{fig:merger_tree_acc}). 
\end{itemize}

When we advance to a more realistic scenario that includes contamination of the in situ population, the performance of the algorithm remains high. The purity and completeness of the recovered clusters decrease only marginally (see Figs.~\ref{fig:P_insitu} and \ref{fig:R_insitu}). The primary challenge introduced by the in situ stars is a reduction in the lookback time for merger detection. The \texttt{HDBSCAN} algorithm confidently identifies only the most recent major accretion events ($z_{\rm acc}<1$), as older, more phase-mixed structures become difficult to distinguish in the chrono-chemodynamical plane (see Fig.~\ref{fig:merger_tree_insitu}).

In conclusion, our work demonstrates that density-based clustering can successfully identify the main merger events of a Milky Way-like galaxy, provided that comprehensive information on its stellar content is available and the algorithm parameters are optimized. The application of this methodology to the Milky Way itself looks encouraging; however, significant challenges remain. Future work must rigorously quantify the performance of the algorithm with incomplete data sets, which reflect the current observational reality, where, for instance, radial velocities or detailed chemical abundances are not available for every star. Furthermore, developing more sophisticated techniques to distinguish the in situ and accreted populations {\it a priori} remains critical to uncover the earliest epoch of Galaxy formation.

\section*{Acknowledgements}

We thank the reviewer for their suggestions and constructing feedback which improved the clarity of the paper. We also thank the \texttt{Auriga} team for providing public access to the particle and merger trees data for their simulations. AS acknowledges a Science Technologies Facilities Council (STFC) Ph.D. studentship at the LIV.INNO Centre for Doctoral Training ``Innovation in Data Intensive Science''. This study used the Prospero high performance computing facility at Liverpool John Moores University.

\section*{Data Availability}

The \texttt{Auriga} simulations can be downloaded through the Globus file transfer service (\url{https://globus.org/}). Detailed instructions on how to access the data are provided in \url{https://wwwmpa.mpa-garching.mpg.de/auriga/data.html}. 

The codes and models developed in the analysis are available under reasonable request to the authors. 



\bibliographystyle{mnras}
\bibliography{example} 





\bsp	
\label{lastpage}
\end{document}